\documentclass[screen]{acmart}

\usepackage{booktabs} % For formal tables
\usepackage{tikz}
\usepackage{booktabs} % For formal tables
\usepackage{float}
\usepackage{multirow}
\usepackage{lipsum}
\usepackage{graphicx}
\usepackage{csquotes}
\usepackage{array}
\usepackage{xcolor}
\usepackage{soul}
\usepackage{multicol}
\usepackage[final]{pdfpages}
\newcolumntype{P}[1]{>{\centering\arraybackslash}p{#1}}
\newcolumntype{M}[1]{>{\centering\arraybackslash}m{#1}}
\usepackage{subfiles}
\usepackage{blindtext}
\usepackage{wrapfig}
\usepackage[utf8]{inputenc}
\usepackage{subcaption}
\usepackage{xfrac}
\usepackage{multicol}
\newcommand{\ignore}[1]{}
\newcolumntype{P}[1]{>{\centering\arraybackslash}p{#1}}
\newcolumntype{M}[1]{>{\centering\arraybackslash}m{#1}}
\AtBeginDocument{%
  \providecommand\BibTeX{{%
    \normalfont B\kern-0.5em{\scshape i\kern-0.25em b}\kern-0.8em\TeX}}}

\begin{document}
\setlength{\tabcolsep}{1.5pt}

\settopmatter{printacmref=false}
\setcopyright{none}
\renewcommand\footnotetextcopyrightpermission[1]{}
\pagestyle{plain}
\title[A Survey on Watching Social Issue Videos]{A Survey on Watching Social Issue Videos among YouTube and TikTok Users}

\author{Shuo Niu}
\email{shniu@clarku.edu}
\orcid{https://orcid.org/0000-0002-8316-4785}
\affiliation{%
  \institution{Clark University}
  \streetaddress{950 Main St.}
  \city{Worcester}
  \state{MA}
  \country{USA}
  \postcode{01610}
}

\author{Dilasha Shrestha}
\email{dshrestha@clarku.edu}
\affiliation{%
  \institution{Clark University}
  \streetaddress{950 Main St.}
  \city{Worcester}
  \state{MA}
  \country{USA}
  \postcode{01610}
}

\author{Abhisan Ghimire}
\email{abghimire@clarku.edu}
\affiliation{%
  \institution{Clark University}
  \streetaddress{950 Main St.}
  \city{Worcester}
  \state{MA}
  \country{USA}
  \postcode{01610}
}

\author{Zhicong Lu}
\email{zhicong.lu@cityu.edu.hk}
% \orcid{https://orcid.org/0000-0002-8316-4785}
\affiliation{%
  \institution{City University of Hong Kong}
  \streetaddress{83 Tat Chee Ave}
  \city{Hong Kong}
%   \state{Massachusetts}
  \country{China}
%   \postcode{01610}
}

\renewcommand{\shortauthors}{Niu, et al.}

\begin{abstract}
The openness and influence of video-sharing platforms (VSPs) such as YouTube and TikTok attracted creators to share videos on various social issues. Although social issue videos (SIVs) affect public opinions and breed misinformation, how VSP users obtain information and interact with SIVs is under-explored. This work surveyed 659 YouTube and 127 TikTok users to understand the motives for consuming SIVs on VSPs. We found that VSP users are primarily motivated by the information and entertainment gratifications to use the platform. VSP users use SIVs for information-seeking purposes and find YouTube and TikTok convenient to interact with SIVs. VSP users moderately watch SIVs for entertainment and inactively engage in social interactions. SIV consumption is associated with information and socialization gratifications of the platform. VSP users appreciate the diversity of information and opinions but would also do their own research and are concerned about the misinformation and echo chamber problems.
 
\end{abstract}


\maketitle

\section{Introduction}
Video-sharing platforms (VSPs) \cite{NiuVSPLR} such as YouTube and TikTok grew into vast online video infrastructures that engage massive users to obtain information, entertain, and socialize with the creators and other viewers. According to a Pew Research survey \cite{perrin_anderson_2019}, YouTube is the most commonly used online platform among US adults in 2020. 81\% of USA adults have used YouTube, up from 73\% in 2019. About half of 18- to 29-year-old adults have used TikTok. The open-access and low barrier of video creation attracted grassroots video creators to share various information and news about social events. About a quarter of US adults get news from YouTube, and people turn to independent channels as often as they use news organization channels \cite{StockingPewResearch}. Online videos may describe subjects negatively and discuss conspiracy theories \cite{StockingPewResearch, MungerSupplyDemand, OttoniRightWingYouTubeChannels}. User-generated videos play a crucial role in obtaining information about social topics. Research on political videos described TikTok creators as performers who externalize opinions via audiovisual performance \cite{SerranoTikTok}. Social interactions with YouTube video creators influence viewers' thoughts toward various social issues \cite{TownerYouTubeElection, BowyerYouTubePolitical}. 
\par
VSPs contrast other networking-based platforms in that social interactions rely on video rather than offline relationships \cite{Burgess2018YouTube:Culture, Hou2018SocialYouTube}. The social and political effects of VSPs have long been recognized in the literature. Researchers have examined the roles of VSPs in discussing social issues such as climate change \cite{Shapiro2014, AllgaierYouTubeClimateChange, HauteaTikTokClimateChange, BaschClimateChangeTikTok}, racial and gender justice \cite{LeeYouTubeControversialTopics, GuoRacismYouTube, HokkaRacismYouTube, KrutrokRacisTikTok}, and public health \cite{BaschYouTubeEbola, LiYouTubeCOVID, BoraYouTubeZika}. Studies also paid attention to fake news and misinformation \cite{ShahidFakeVideos, MitraYouTubeMisinformation} and content moderation \cite{JhaverYouTubeCommentModeration} on VSPs. Social issue videos (SIVs) present or discuss various social matters involving controversy or uncertainty over the well-being of a large number of people \cite{RappoportSocialIssue}. Video creators on YouTube differentiate themselves from mainstream media and claim to be alternative but more creditable sources \cite{LewisYouTubeMicrocelebrity}. Meanwhile, SIVs may contain diverse views ranging from dystopian to utopian \cite{HerrmanTikTok}. Extremist creators could supply ideological content to meet viewers' demand for extreme content \cite{MungerSupplyDemand}. However, a knowledge gap exists in understanding VSP users' demand and consumption of social issue information. Research on the role of social media videos in society could benefit from the knowledge of how VSP users seek information, gain entertainment, and socialize with others through SIVs. 
\par
In this paper, we present a survey of how US users of YouTube and TikTok -- the two largest video-sharing platforms -- watch social-issue-related videos. This study examines the uses and gratifications of SIVs on the two platforms. We surveyed 659 YouTube users and 127 TikTok users in the USA to describe their motivation for using YouTube and TikTok in general and watching user-generated videos to understand social issues. We also compare users of the two groups in SIV watching. The uses and gratifications theory is a widely used theoretical instrument to measure social media motivations. In a nutshell, this work addresses four research questions:
% YouTube is the largest and most highly visited online video-sharing service that features user-generated content and community activities \cite{Burgess2018YouTube:Culture}. TikTok is a newer VSP that features short videos for mobile devices, attracting young users to engage in video interactions \cite{MontagTikTokUse}. Studies examining the motivations for using YouTube and TikTok suggested both platforms offer informational, entertainment, and social gratification \cite{KhanSocialMediaEngagement, BufYouTube}. YouTube contains more educational and specialized videos \cite{MaroofYouTubeTikTok}; while TikTok usage is driven by archiving, self-expression, social interaction, and escapes \cite{OmarTikTok}. 
\begin{itemize}
    \item RQ1: Why do users use YouTube and TikTok and what gratifications do they receive from video-sharing platforms?
    \item RQ2: How do video-sharing platform users watch social issue videos to obtain information, entertainment, and socialization gratifications?
    \item RQ3: How do gratifications with video-sharing platforms affect watching social issue videos?
    \item RQ4: How do social issue videos affect users' understanding, opinion, and action?
\end{itemize}
RQ1 examines the uses and gratifications of YouTube and TikTok to identify the general user motives. RQ2 asks questions on three gratification themes to investigate how YouTube and TikTok users watch SIVs. RQ3 consists of association analysis to examine what platform motives predict information, entertainment, and socialization gratifications with SIVs. RQ4 incorporates an open-ended question to probe viewers' thoughts on the benefits and problems with SIVs on VSPs. VSP users are primarily attracted by information and entertainment gratifications. Regarding SIVs, YouTube users are slightly negative about obtaining social issue information from videos while TikTok users are more likely to do so. Users of both platforms agree that watching SIVs could affect opinions, while they would do their own research to check correctness. VSP users moderately use SIVs for entertainment and do not tend to engage in social activities around SIVs. Compared to YouTube users, TikTok users are more likely to follow SIV creators and keep up with new videos. Watching and interacting with SIVs are associated with information and socialization gratifications. Users watch SIVs to learn new information, know different opinions, and shape their views. However, VSP users are also concerned about biased and misleading information, negative feelings, and echo chamber problems. 

\section{Related Work}

\subsection{Social Media and Video-Sharing Platforms}
YouTube and TikTok are known as places to create and circulate user-generated and personally meaningful videos \cite{YouTubeParticipatoryCulture}. Two core cultural logics of YouTube -- diverse participation and celebrity making -- center YouTube platform ecology \cite{Burgess2018YouTube:Culture}. The enormous number of YouTubers and TikTokers create rich information and entertaining content for the viewers; meanwhile, they socialize with the viewers to form online communities \cite{PreeceWeTube, LuTikTok}. In contrast to platforms based on friending and networking, VSP users rely on video creation rather than offline relationships \cite{Burgess2018YouTube:Culture}. 
\par
\par
YouTube and TikTok are popular video-sharing platforms, but some key differences exist. YouTube is a general video-sharing platform where users can upload a wide range of videos, including vlogs, tutorials, music videos, and more \cite{Che2015ACharacteristics}. TikTok, on the other hand, is primarily focused on short-form, lip-sync, dance, and comedy videos \cite{Shutsko2020TikTok}. YouTube allows users to upload videos of any length. The average length of YouTube videos is 11.7 minutes\footnote{https://www.statista.com/statistics/1026923/youtube-video-category-average-length/}, while the optimal video length of TikTok videos is between 21 and 34 seconds\footnote{https://www.wired.com/story/tiktok-wants-longer-videos-like-not/}. YouTube caters to a more diverse and older demographic, while TikTok is primarily popular among younger users \cite{perrin_anderson_2019}. YouTube and TikTok provide several monetization options for content creators, including advertising revenue and sponsorship \cite{YouTube5Year, Zhang2021TikTok}. TikTok's algorithm is designed to show users videos they're likely to engage with \cite{Zhang2021TikTokAlgorithm}. YouTube's algorithm considers a wider range of factors, including keywords, engagement, and user behavior \cite{CovingtonYouTubeRecommendations}. 

\subsection{Research on Social Issue Videos on Video-Sharing Platforms}
Researchers have paid attention to the socio-technical impact of online videos on public opinions and collective actions. Studies have examined how video creators offer general information about climate change and environmental justice education \cite{Shapiro2014, AllgaierYouTubeClimateChange, HauteaTikTokClimateChange, BaschClimateChangeTikTok, NiuTeamTrees}, embed perpetuated stereotypes and racism ideologies in everyday videos \cite{LeeYouTubeControversialTopics, GuoRacismYouTube, HokkaRacismYouTube, KrutrokRacisTikTok}, and share information about health emergencies such as Ebola, Zika, and COVID-19 \cite{BaschYouTubeEbola, LiYouTubeCOVID, BoraYouTubeZika}. Although SIVs on VSPs amplify discussion, a considerable proportion of online videos contain misinformation \cite{BoraYouTubeZika, BaschClimateChangeTikTok}, conspiracy theories \cite{AllgaierYouTubeClimateChange, HerrmanTikTok, TangMisinformation}, and biased and even radicalized opinions \cite{LeeYouTubeControversialTopics, GuoRacismYouTube, KrutrokRacisTikTok}. 
\par
Researchers have examined the informational, entertaining, and social affordances of VSP in spreading social issue knowledge. For example, Video creators advocating social changes present and understand themselves as agencies and choice-makers as part of a supportive online community \cite{RabyYouTubeSocialChange}. YouTubers act as micro-celebrities among viewers' communities, conflate the mainstream media, and stress their relatability, authenticity, and accountability \cite{LewisYouTubeMicrocelebrity}. Audiences who demand radicalized content exist, and creators serve as suppliers of radical ideologies videos \cite{MungerSupplyDemand}. TikTokers tend to show entertaining and playful political participation \cite{VijayTikTokPolitics}. Information falsehood on TikTok can be spread in entertainment forms like humorous memes \cite{BaschCOVIDTikTok} and dance challenges \cite{LiCOVIDTikTok}. Adolescent TikTokers' video creation is affected by the popular trends on the platform but may engage with harmful content such as racism \cite{LiuTikTokRacistTrend}. However, only a few studies explored how YouTube and TikTok users consume SIVs; the research on SIVs is behind VSPs' virality among the public \cite{WeimannSpreadingHateTikTok}. 
\par
There is emerging research on the roles of VSPs in various social topics in social media research. One realm of research seeks to understand the creator-viewer interactions and community activities during social events \cite{NiuTeamTrees, RohdeSyriaVideo}. Another realm examined how the technology designs on VSPs such as recommendation algorithms \cite{MitraYouTubeMisinformation, BuntainYouTubeRecommendation}, commenting tools \cite{HeChinaOnlineCrisisDanmaku}, and moderation techniques \cite{JhaverYouTubeCommentModeration}. However, previous research mostly focuses on understanding the video content related to social issues; there lacks a ground understanding of video consumption from viewers' perspectives. Researchers need to examine the differences between a general VSP like YouTube and a short-video platform like TikTok. This understanding will help platform designers and policymakers to understand why VSP users interact with SIVs, thus designing platform features to promote the positive use of VSPs for consuming social issue information.

\subsection{The Uses and Gratifications of YouTube and TikTok}
Prior studies examined viewers' motivations for using VSPs with the uses and gratifications theory \cite{KatzUGT, WhitingUseGratification}. The uses and gratifications theory explains why people use media, what they use them for, and how users deliberately choose media \cite{KatzUGT}. Ten uses and gratifications for social media were identified \cite{WhitingUseGratification}: social interaction, information seeking, passing the time, entertainment, relaxation, expression of opinions, communicatory utility, convenience utility, information sharing, and knowledge about others. 
\par
Researchers have examined the usage of YouTube and TikTok with the uses and gratifications theory. YouTube viewers satisfy users' need for information and relaxation, while content creators have higher demands for social recognition \cite{BufYouTube}. Higher socialization and entertainment motivations indicate higher involvement and compulsive usage while browsing information is irrelevant to the level of YouTube involvement \cite{WangUGTYouTube, KlobasYouTube, BalakrishnanYouTube}. YouTube users who are gratified by the relaxation and entertainment motives tend to like or dislike the videos \cite{KhanSocialMediaEngagement}. YouTube videos about political and social topics are perceived as more entertaining, emotional, motivating, and more subjective and manipulating \cite{ZimmermannYouTube}. TikTok's ``\textit{For You}'' page sends extreme content even if the users do not follow such creators \cite{WeimannSpreadingHateTikTok}. But similar to YouTube, TikTok users primarily use the platform for archiving, self-expression, social interaction, and escapism \cite{OmarTikTok, MengTikTok}. Entertainment gratification is the dominant motivation using TikTok \cite{BossenUGTikTok, AhlseTikTok, LuTikTok}, while socialization gratification is connected to more intensive parasocial relationships with the video creators \cite{YangTikTokChina}. People may avoid TikTok for fear of addiction or the perception of low-quality content \cite{LuTikTokNonUse}. YouTube and TikTok offer different video-watching experiences. However, it is unknown how VSP users perceive the use and gratification of SIVs on YouTube and TikTok.

\section{Uses and Gratifications Theory}
We adopt Whiting's framework \cite{WhitingUseGratification} of uses and gratifications of social media as a theoretical framework to design survey questions. This theory includes ten themes that reveal common motivations for using social media (see Table \ref{tab:framework}). Based on the ten themes, we identify three main types of gratifications with video-sharing platforms. \textit{Information gratification} considers how viewers interact with the in-video information. \textit{Entertainment gratification} refers to motivations of using videos for entertainment and enjoyment, passing the time when feeling bored, and relaxation and relieving day-to-day stress. \textit{Socialization gratification} is about the social interactions with the video creators and other viewers, expressing thoughts and opinions through videos, sharing information about oneself, and communicating with others. 
\par
\begin{table*}[!ht]
    \centering
    \scalebox{0.75}{
        \begin{tabular}{|P{3cm}|p{4cm}|p{11cm}|}
            \hline
            U\&G & theme & definition \\
            \hline
            \multirow{3}{2.1cm}{Information} & Information seeking & Use video-sharing platforms to seek out information or to self-educate\\
            \cline{2-3}
            & Convenient utility & Provide convenience or usefulness to find information\\
            \cline{2-3}
            & Knowledge about others & Use video-sharing platforms to know about others' lives, opinions, or attitudes\\
            \hline
            \multirow{3}{2.1cm}{Entertainment} & Entertainment & Use video-sharing platforms for entertainment and enjoyment\\
            \cline{2-3}
            & Pass time & Use video-sharing platforms to occupy time and relieve boredom\\
            \cline{2-3}
            & Relaxation & Use video-sharing platforms to relieve day-to-day stress\\
            \hline
            \multirow{4}{2.1cm}{Socialization} & Social interaction & Use video-sharing platforms to communicate and interact with others\\ 
            \cline{2-3}
            & Express opinion & Use video-sharing platforms to express thoughts and opinions\\
            \cline{2-3}
            & Information sharing & Use video-sharing platforms to share information about oneself with others\\
            \cline{2-3}
            & Communicatory utility & Facilitate communication and provide information to share with others\\
            \hline
        \end{tabular}
    }
    \caption{The survey question themes and definitions based on the uses and gratifications theory}
    \label{tab:framework}
\end{table*}
RQ1 examines the uses and gratifications of YouTube and TikTok. Questions for RQ1 do not particularly ask about SIVs but about viewers' general motivations when interacting with the platform. RQ2 pertains to SIVs and includes survey questions that probe how YouTube and TikTok users watch SIVs to obtain information, entertain themselves, and socialize. RQ3 explores whether the motivations for using VSPs affect the consumption of SIVs. We build linear regression models to verify the relationships between users' VSP usage motivations and the informational, entertainment, and social activities with SIVs. RQ4 queries how SIVs affect users through an open-ended question.

\section{Survey Design}
The participants were recruited from Amazon Mechanical Turk (MTurk). This study only recruited participants in the USA because participants on MTurk are international, and people from different countries may have different understandings of various social issues. Participants must be at least 18 years old to take the survey. A study found MTurk workers are active on social media and have more experience contributing to online content \cite{ShawMTurk}. Therefore, participants on MTurk can represent active social media users and could be affected by SIVs. To ensure the answer quality, the participants must complete at least 5,000 tasks with an approval rate higher than 97\%. Two simple math questions were incorporated as an attention test. If the participants failed the attention test, their response was rejected. 
\par
Many users use both YouTube and TikTok, but we asked each participant to recognize themselves as a primary YouTube user or TikTok user to obtain their opinion about their preferred VSP. Participants could take only one of the YouTube or TikTok versions of the questionnaire to be surveyed about their primary VSP. In the beginning, a multiple-choice question asked the participants whether they used YouTube more often, TikTok more often, or neither of the platforms. Two versions of the survey were prepared with questions about either YouTube or TikTok. If the participant selected YouTube (TikTok), we categorized the participant as a YouTube (TikTok) user and asked the participant to fill out the YouTube (TikTok) version of the survey. The participant will not proceed if they choose neither. The questionnaire included demographic questions asking participants' age, gender, and ethnicity. We also asked how frequently the users used the VSP during the past month. The following statistical analysis includes the differences in user demographics and uses frequency as independent variables.
\par
To address RQ1, ten questions asked the participants to rate the importance of each motive specified by the uses and gratifications theory \cite{WhitingUseGratification}. The questions were based on the definitions specified in Table \ref{tab:framework}. The questions were 7-point Likert style with options from ``not at all important'' to ``very important.'' RQ2 focuses on participants' experiences with SIVs on VSPs (Table \ref{tab:social_issue_questions}). The questionnaire presented a list of 37 common social issue topics\footnote{https://educationforjustice.org/topics/social-justice-topics/} before the questions about SIVs to give participants examples of social issues. We also gave examples of social issues (e.g., voting rights, climate justice, health care, refugee crisis, etc.) in every question where we refer to ``social issue videos.'' To standardize the answers, all questions were either 7-point Likert-style multi-choice questions (e.g., ``very rarely'', ``rarely'', ``somewhat rarely'', ``sometimes'', ``somewhat often'', ``often'', and ``very often'') or questions with options of seven different levels. I1 to I5 examined the information-seeking theme. I6 to I10 were questions about the convenient utility theme. I11 to I13 were questions regarding knowledge about others. E1 to E4 were under entertainment themes. Social themes were examined by questions S1 to S7. See Table \ref{tab:social_issue_questions} for the questions. RQ4 probed how SIVs affect users with an open-ended question asking the participants to write a short answer reflecting how SIVs affect opinions, understandings, and actions. This question required participants to respond with at least 50 characters.
\par
\begin{table*}[!h]
    \centering
    \scalebox{0.75}{
        \begin{tabular}{|P{1.5cm}|p{14cm}|p{3cm}|}
            \hline
            & Uses of Video-Sharing Platforms & Gratifications \\
            \hline
            I1 & Obtain information related to current social issues & \multirow{6}{3cm}{Information seeking} \\
            I2 & When you have questions related to social issues, use YouTube/TikTok to seek answers or solutions & \\
            I3 & Trust the information you received through YouTube/TikTok videos & \\
            I4 & Do research on other sources to figure out if the information is correct & \\
            I5 & Affect your opinions about social issues & \\
            \hline
            I6 & Watch recommended social issue videos & \multirow{5}{3cm}{Convenient utility}\\
            I7 & Search social issue videos you want to watch & \\
            I8 & Upload social issue videos & \\
            I9 & Comment on social issue videos and interact with other viewers & \\
            I10 & Share social issue videos to other social media & \\
            \hline
            I11 & Know other people's opinions on social issues & \multirow{3}{3cm}{Knowledge about others}\\
            I12 & How many YouTubers/TikTokers who posted videos about social issues do you follow? & \\
            I13 & How often do you keep up with the new social issue videos of YouTubers/TikTokers you follow? & \\
            \hline
            E1 & Watch social issue videos for entertainment & Entertainment\\
            \hline
            E2 & Watch social issue videos when you have idle time or when you feel bored & Pass time \\
            \hline
            E3 & Watch social issue videos help you feel relaxed & \multirow{2}{3cm}{Relaxation}\\
            E4 & Watch social issue videos make you feel stressful & \\
            \hline
            S1 & Contact the video creators through comments & \multirow{3}{3cm}{Social interaction}\\
            S2 & Contact the video creators through direct messages or social media & \\
            S3 & Contact other viewers through comments or direct messages & \\
            \hline
            S4 & Have you ever uploaded videos to YouTube/TikTok to express thoughts and opinions related to social issues? & \multirow{2}{3cm}{Express and share opinions}\\
            S5 & How often do you leave comments about social issues to express thoughts and opinions & \\
            \hline
            S6 & Share YouTube/TikTok videos related to social issues to other social media & \multirow{2}{3cm}{Communicatory utility}\\
            S7 & Share YouTube/TikTok videos in life with your family and friends & \\
            \hline
        \end{tabular}
    }
    \caption{Survey questions about watching social issue videos on YouTube and TikTok}
    \label{tab:social_issue_questions}
\end{table*}

\section{Analysis Method}
Least-square regression (LSR) models were built to predict the answers to each question in RQ1 and RQ2. In each LSR model, the independent variables included the platform (YouTube or TikTok), age (an ordinal factor with seven levels), gender (a nominal factor with three categories), ethnicity (a nominal factor with eight categories), and frequency (an ordinal factor with seven levels) as simultaneous independent factors. The dependent variables were the scores of 7-point Likert-style or 7-level answers. The alpha value to determine significance is set to 0.05 with the Bonferroni adjustment. Considering unequal numbers of participants of the two platforms exist, we perform posthoc analysis with Welch's test to justify the difference between the unbalanced groups. This study focuses on understanding YouTube and TikTok usage; we do not discuss the effects of user age, gender, ethnicity, and platform use frequency to avoid losing focus. Also, the discussion cannot be comprehensive without enough participants of some age (e.g., non-binary participants), ethnic (e.g., some minority groups), and use-frequency groups (e.g., users who use less than an hour every month).
\par
RQ3 examines whether users' information-seeking, entertainment, and socialization with SIVs are associated with the three types of gratifications. LSR models used the average scores of RQ1 as independent variables to predict participants' ratings of SIV watching in RQ2. We chose a subset of RQ2 questions to represent information, entertainment, and social gratifications with SIVs. We selected information-seeking (average ratings of I1 and I2), entertainment (E1), and social interactions (average ratings of S1, S2, and S3) scores as dependent variables. 
%Seeking social issue information is the average of I1 and I2. Entertainment with SIVs is the rating of E1. Social interaction is reflected by the mean rating of S1, S2, and S3.
\par
The thematic analysis method \cite{BraunThematicAnalysis} was used to derive themes from participants' open-ended responses to RQ4. We performed the initial code generation, theme searching, theme reviewing and agreement analysis, and data encoding to process the answers. One hundred randomly selected responses were used to generate the initial codebook, and another 100 were encoded with the codebook for theme searching and refining. Then two authors performed two rounds of annotating 50 responses to validate the codebook further, clarify the definitions, and calculate the agreement level between the raters. The agreement level (Cohen's Kappa) raised from 0.62 in the first round to 0.72 in the second round. Since the agreement reached a substantial level, the two authors split the rest of the responses and annotated them with the codebook. We dropped irrelevant responses such as random texts, non-understandable sentences, and discussing other media (e.g., TV news) from the data. We partially or fully removed 42 (6.37\%) responses from YouTube users and 19 (14.96\%) responses from TikTok users.

\section{Survey Participants}
The survey had 786 participants with completed answers, including 659 (83.84\%) who identified themselves as YouTube users and 127 (16.16\%) who identified as TikTok users. Participants' age group is converted by a ranking scale of 1 for ``18-24'' and 7 for ``75-84'' (no participant selected 85 or above). Unsurprisingly, the Wilcoxon test suggests TikTok users are significantly younger than YouTube users ($p=0.0029$, $\chi^2=8.87$). The average age ranking is 2.69 ($SD=1.17$) for TikTok users and 2.99 ($SD=1.14$) for YouTube users (two participants didn't provide their age group). Pearson's test suggests participants from the two platforms have different gender distributions ($p<0.0001$, $\chi^2=0.0001$). The TikTok group has significantly more female-identified participants than the YouTube group (not enough participants to compare non-binary participants). There is no significant difference in ethnicity and usage frequency between the participants of the two groups. 
\par
\begin{figure}[!h]
\centering
\scalebox{0.47}{
    \begin{tabular}{ll}
        
        \begin{subfigure}{\linewidth}
        \includegraphics[width=\linewidth]{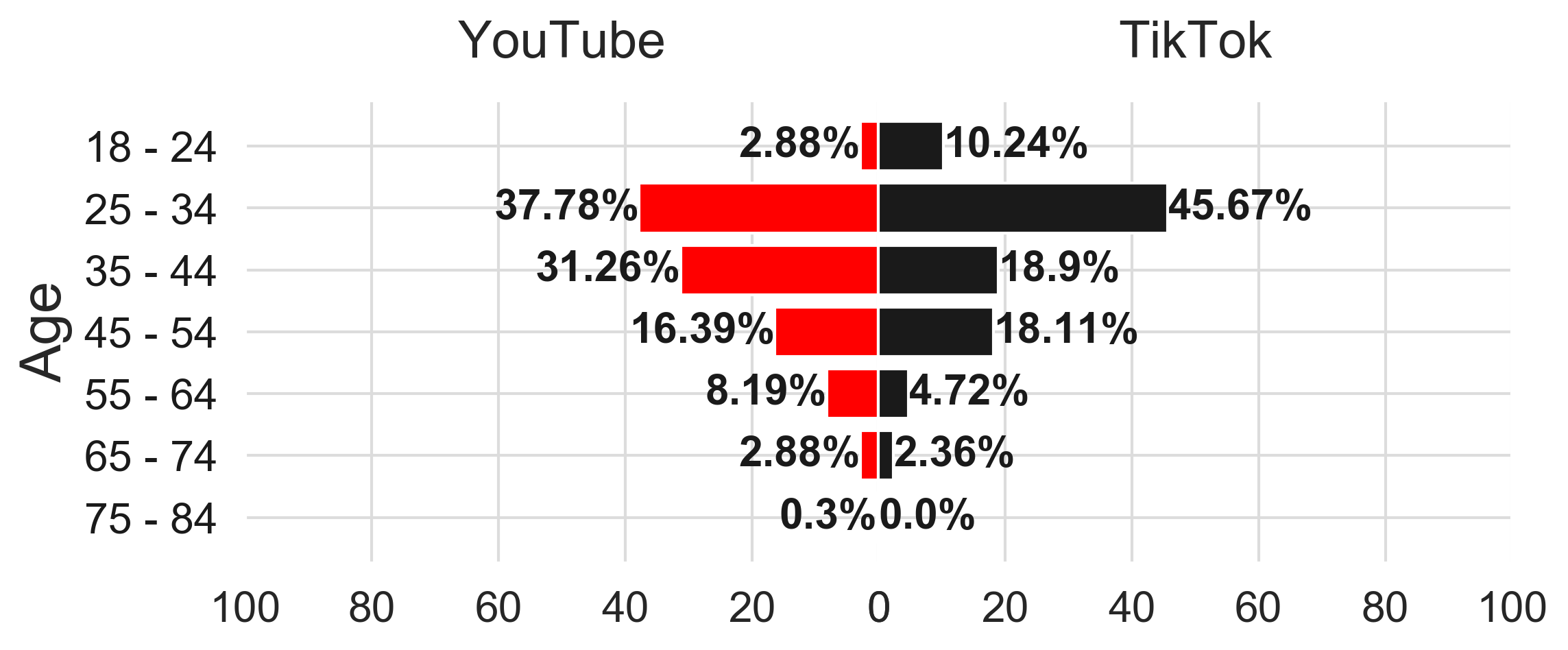}
        \end{subfigure}
        & 
        \begin{subfigure}{\linewidth}
        \includegraphics[width=\linewidth]{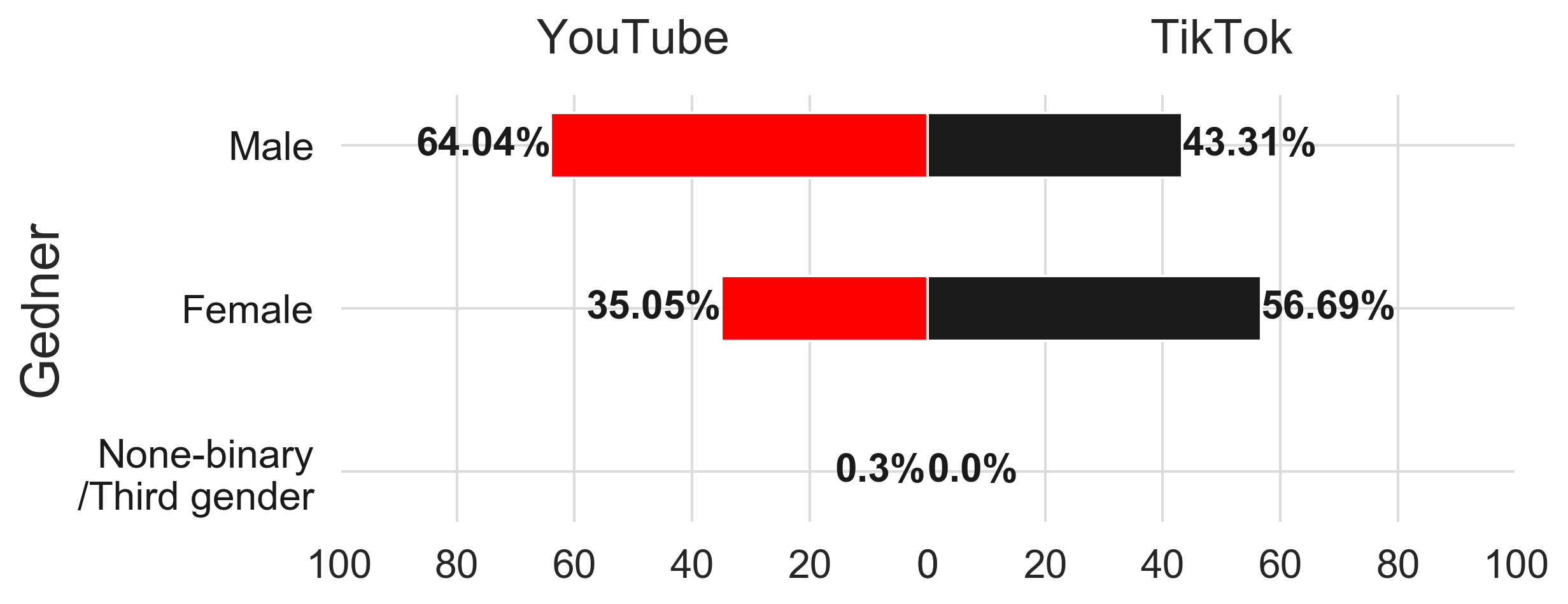}
        \end{subfigure}
        \\ 
        \begin{subfigure}{\linewidth}
        \includegraphics[width=\linewidth]{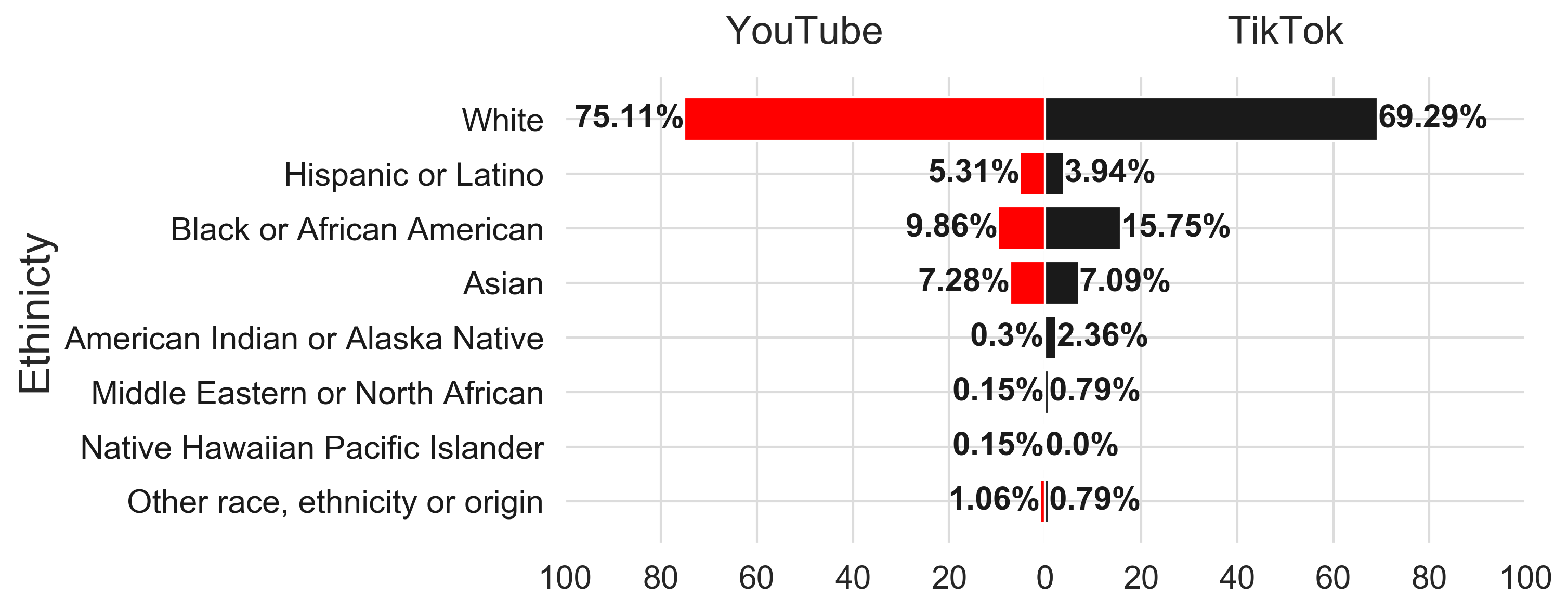}
        \end{subfigure}
        & 
        \begin{subfigure}{\linewidth}
        \includegraphics[width=\linewidth]{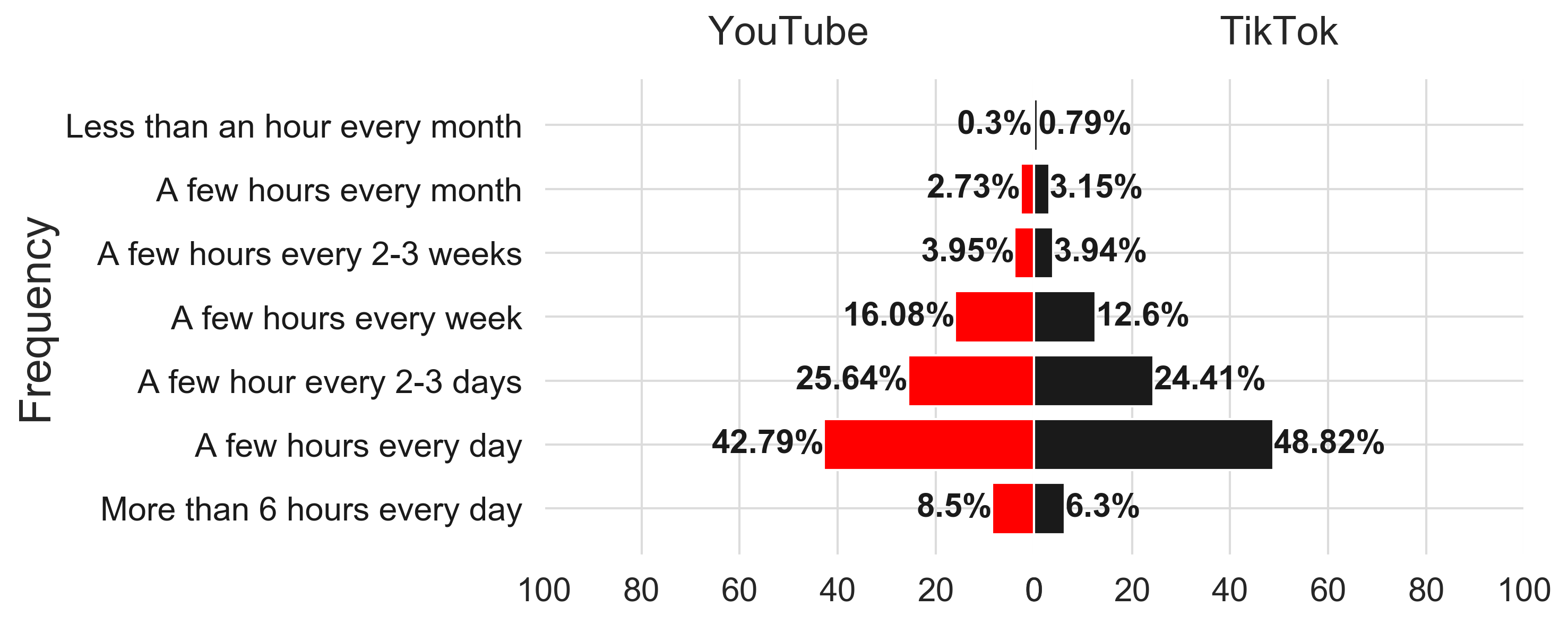}
        \end{subfigure}
    \end{tabular}
}
\caption{The distribution of age, gender, ethnicity, and use frequency of YouTube and TikTok participants}
\label{fig:demographics}
\end{figure}

\section{Results}
\subsection{RQ1: The Differences in the Gratifications Between YouTube and TikTok Users}
\subsubsection{Information Gratifications}

\begin{figure}[!h]
    \centering
    \begin{tabular}{c}
        \scalebox{0.96}{
            \includegraphics[width=\linewidth]{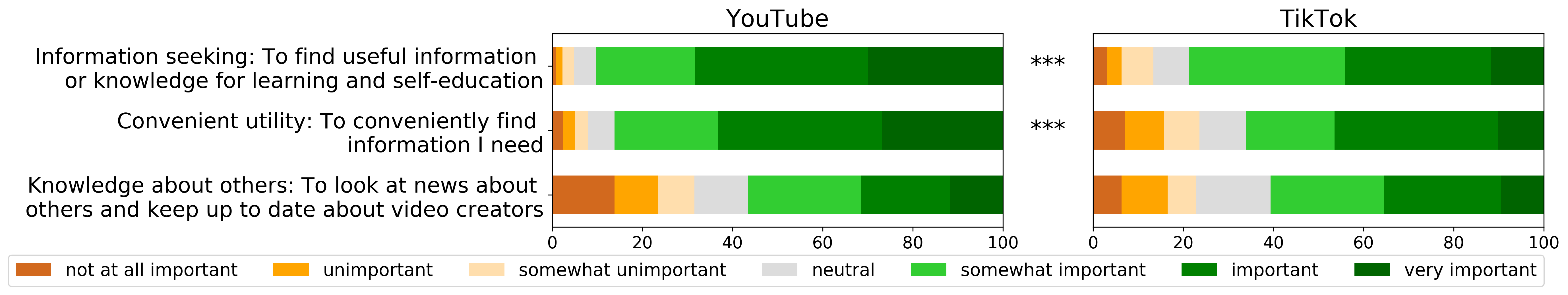}
        }\\
        \scalebox{0.7}{
            \begin{tabular}{p{2cm}p{0.15\linewidth}p{0.15\linewidth}|p{0.15\linewidth}p{0.15\linewidth}|p{0.15\linewidth}p{0.15\linewidth}}
                \toprule
                &\multicolumn{2}{l}{\textbf{Information seeking}}&\multicolumn{2}{l}{\textbf{Convenient utility}}&\multicolumn{2}{l}{\textbf{Knowledge about others}}\\
                & \multicolumn{2}{l}{$F=3.54$,  $r^2=0.09$,  $p<0.0001$} & \multicolumn{2}{l}{$F=3.62$,  $r^2=0.10$,  $p<0.0001$} & \multicolumn{2}{l}{$F=7.29$,  $r^2=0.17$,  $p<0.0001$}\\
                \midrule
                factor & $F$ ratio & $p$ & $F$ ratio & $p$ & $F$ ratio & $p$ \\
                Platform & 36.36 & $<$0.0001 & 38.44 & $<$0.0001 & - & -\\
                Age & - & - & - & - & - & - \\
                Gender & - & - & - & - & - & - \\
                Ethnicity & - & - & - & - & 2.74 & 0.0081 \\
                Frequency & - & - & 4.03 & 0.0005 & 18.66 & $<$0.0001\\
            \bottomrule
            \end{tabular}
    }
    \end{tabular}
    \caption{The distribution and statistic results of ratings for information seeking, convenient utility, and knowledge about others. Only statistics with a significant p-value are presented in the table. $p*<0.05$, $p**0.01$, $p***<0.001$. Adjusted $\alpha=0.0167$.}
    \label{fig:motivation_information}
\end{figure}

The participants rated information needs as important gratifications for using VSPs (Figure \ref{fig:motivation_information}). The questions that probe the importance of \textit{information seeking}, \textit{convenient utility}, and \textit{knowledge about others} (1-not at all important, 7-very important) received average ratings of 5.69 ($SD=1.23$), 5.47 ($SD=1.47$), and 4.36 ($SD=1.89$), respectively. When asked to what degree the participants are motivated by finding valuable information or knowledge for learning and self-education, YouTube users rated 5.80 ($SD=1.16$) and TikTok rated 5.42 ($SD=1.40$). YouTube users rated convenient utility an average of 5.61 ($SD=1.37$) and TikTok users rated it an average of 4.76 ($1.75$). The LSR models that predict information seeking and convenient utility by the five independent factors suggest platform is a significant factor (Figure \ref{fig:motivation_information}). Posthoc with Welch's test suggests a significant difference exists (both $p_{Welch}<0.0001$). YouTube users rated information-seeking and convenient utility higher than TikTok users. These results suggest that both YouTube and TikTok are used to satisfy information needs. But YouTube participants are more motivated by information seeking and convenient utility than TikTok users. 
\par

\subsubsection{Entertainment Gratifications}
\begin{figure}[!h]
    \centering
    
    \begin{tabular}{c}
        \scalebox{0.96}{
            \includegraphics[width=\linewidth]{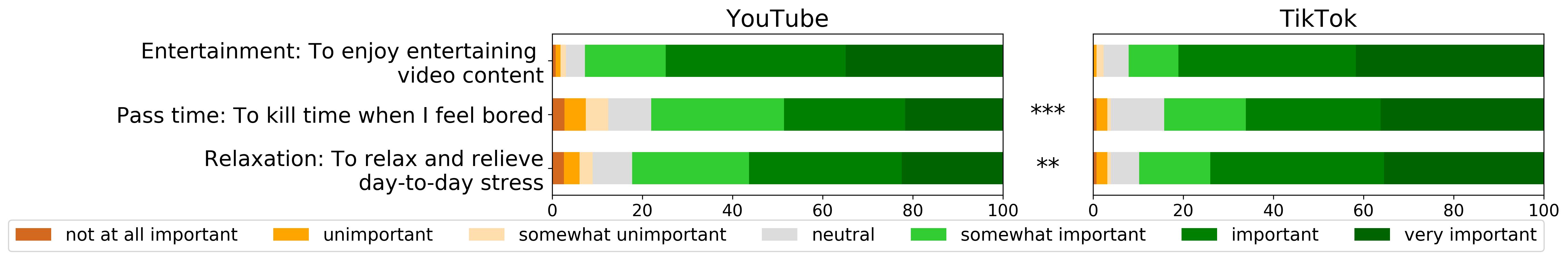}
        }\\
        \scalebox{0.7}{
            \begin{tabular}{p{2cm}p{0.15\linewidth}p{0.15\linewidth}|p{0.15\linewidth}p{0.15\linewidth}|p{0.15\linewidth}p{0.15\linewidth}}
            \toprule
                &\multicolumn{2}{l}{\textbf{Entertainment}}&\multicolumn{2}{l}{\textbf{Pass time}}&\multicolumn{2}{l}{\textbf{Relaxation}}\\
                & \multicolumn{2}{l}{$F=3.32$,  $r^2=0.09$,  $p<0.0001$} & \multicolumn{2}{l}{$F=5.59$,  $r^2=0.14$,  $p<0.0001$} & \multicolumn{2}{l}{$F=4.51$,  $r^2=0.12$,  $p<0.0001$}\\
                \midrule
                factor & $F$ ratio & $p$ & $F$ ratio & $p$ & $F$ ratio & $p$ \\
                Platform & - & - & 13.51 & 0.0003 & 7.56 & 0.0060 \\
                Age & 3.35 & 0.0029 & 9.98 & $<$0.0001 & 3.65 & 0.0014 \\
                Gender & - & - & - & - & 6.13 & 0.0023 \\
                Ethnicity & - & - & 2.81 & 0.0068 & - & - \\
                Frequency & 6.78 & $<$0.0001 & 3.46 & 0.0022 & 7.46 & $<$0.0001 \\
            \bottomrule
            \end{tabular}
        }
    \end{tabular}
    \caption{The distribution and statistic results of ratings for entertainment, pass the time, and relaxation. Only statistics with a significant p-value are presented in the table. Adjusted $\alpha=0.0167$.}
    \label{fig:motivation_relaxation}
\end{figure}

Entertainment is a highly rated gratification for using VSPs. All three questions about entertainment gratifications received positive ratings (1-not at all important, 7-very important, Figure \ref{fig:motivation_relaxation}). When asked about the degree to which VSPs are used to enjoy entertaining content, YouTube and TikTok users rated 6.12 ($SD=1.00$) and 5.97 ($SD=1.08$). Similarly, killing time when feel bored ($MEAN_{YouTube}=5.26$, $SD=1.49$, $MEAN_{TikTok}=5.79$, $SD=1.28$) and relaxation and relieving stress ($MEAN_{YouTube}=5.43$, $SD=1.41$, $MEAN_{TikTok}=5.91$, $SD=1.19$) were also rated as important gratifications. The LSR models suggest platform is a significant predictor for passing the time and relaxation (Figure \ref{fig:motivation_relaxation}). TikTok users rated significantly higher on the passing time (posthoc $p_{Welch}<0.0001$) and relaxation questions (posthoc $p_{Welch}<0.0001$). It indicates that users interact with VSP videos for entertainment, while TikTok users are more likely to occupy idle time and relieve day-to-day stress than YouTube users.
\par

\subsubsection{Social Gratifications}
\begin{figure}[!h]
    \centering
    \begin{tabular}{c}
        \scalebox{0.96}{
            \includegraphics[width=\linewidth]{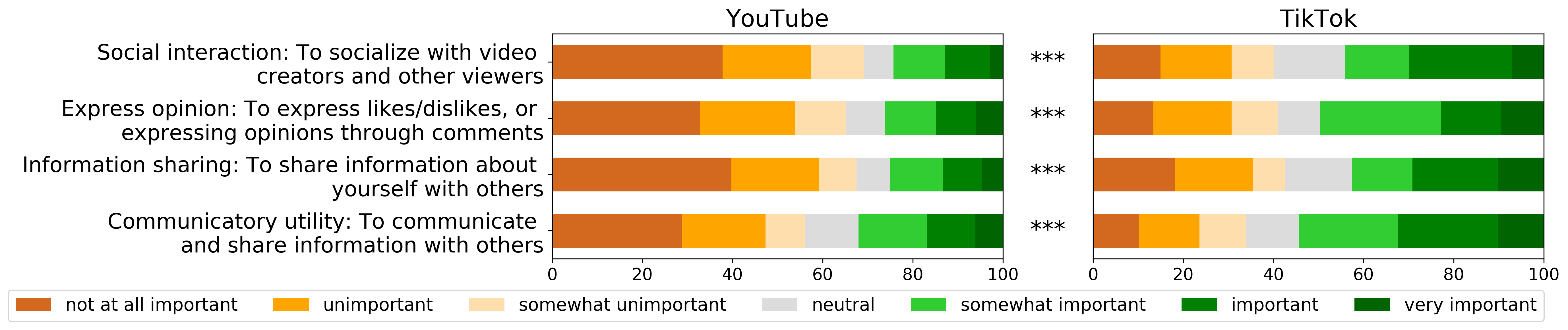}
        }\\
        \scalebox{0.7}{
        \begin{tabular}{p{1.5cm}p{0.15\linewidth}p{0.15\linewidth}|p{0.15\linewidth}p{0.15\linewidth}|p{0.15\linewidth}p{0.15\linewidth}|p{0.15\linewidth}p{0.15\linewidth}}
        \toprule
            &\multicolumn{2}{l}{\textbf{Social interaction}}&\multicolumn{2}{l}{\textbf{Express opinion}}&\multicolumn{2}{l}{\textbf{Information sharing}}&\multicolumn{2}{l}{\textbf{Communicatory utility}}\\
            & \multicolumn{2}{l}{$F=8.12$,  $r^2=0.19$,  $p<0.0001$} & \multicolumn{2}{l}{$F=6.37$,  $r^2=0.16$,  $p<0.0001$} & \multicolumn{2}{l}{$F=6.86$,  $r^2=0.17$,  $p<0.0001$} & \multicolumn{2}{l}{$F=6.22$,  $r^2=0.15$,  $p<0.0001$}\\
            \midrule
            factor & $F$ ratio & $p$ & $F$ ratio & $p$ & $F$ ratio & $p$ & $F$ ratio & $p$ \\
            Platform & 35.85 & $<$0.0001 & 22.84 & $<$0.0001 & 25.82 & $<$0.0001 & 24.40 & $<$0.0001 \\
            Age & - & - & - & - & - & - & - & -\\
            Gender & - & - & - & - & - & - & - & -\\
            Ethnicity & 4.56 & $<$0.0001 & 2.66 & 0.0100 & 2.84 & 0.0062 & - & - \\
            Frequency & 15.16 & $<$0.0001 & 13.20 & $<$0.0001 & 12.93 & $<$0.0001 & 13.38 & $<$0.0001 \\
        \bottomrule
        \end{tabular}
    }
    \end{tabular}
    \caption{The distribution and statistic results of ratings for social interaction, express opinion, information sharing, and communicatory utility ratings. Only statistics with a significant p-value are presented in the table. $p*<0.05$, $p**<0.01$, $p***<0.001$. Adjusted $\alpha=0.0125$.}
    \label{fig:motivation_social}
\end{figure}

Socialization is not an important gratification for VSP users (Figure \ref{fig:motivation_social}). When asked whether they use VSPs to socialize with creators and other viewers (1-not at all important, 7-very important), YouTube users rated 2.76 ($SD=1.88$), and TikTok users rated 3.95 ($SD=1.95$). Participants were also less motivated by using VSPs to express likes/dislikes and opinions ($MEAN_{YouTube}=2.95$, $SD=1.95$, $MEAN_{TikTok}=3.97$, $SD=1.92$). YouTube users also tend not to share information about themselves with others ($MEAN_{YouTube}=2.77$, $SD=1.95$, $MEAN_{TikTok}=3.86$, $SD=2.05$). The only positive rating in the socialization gratifications was that TikTok users rated it important to communicate with others ($MEAN_{TikTok}=4.29$, $SD=1.88$). In contrast, YouTube users did not find this gratification important ($MEAN_{TikTok}=3.23$, $SD=1.99$). The LSR model suggests a significant difference between the two platforms (Figure \ref{fig:motivation_social}). TikTok users rated significantly higher than YouTube users for all four questions (all posthoc $p_{Welch}<0.0001$). Socializing with creators and other viewers is not a key motivator. But TikTok users are more motivated by social interaction, expressing opinions, sharing information, and communicatory utility than YouTube users. 
\par

\subsection{RQ2: Watching Social Issue Videos on Video-Sharing Platforms}

\subsubsection{Information Gratification and Social Issue Videos}
\begin{figure}[!h]
    \centering
    \begin{tabular}{c}
        \scalebox{0.96}{
            \includegraphics[width=\linewidth]{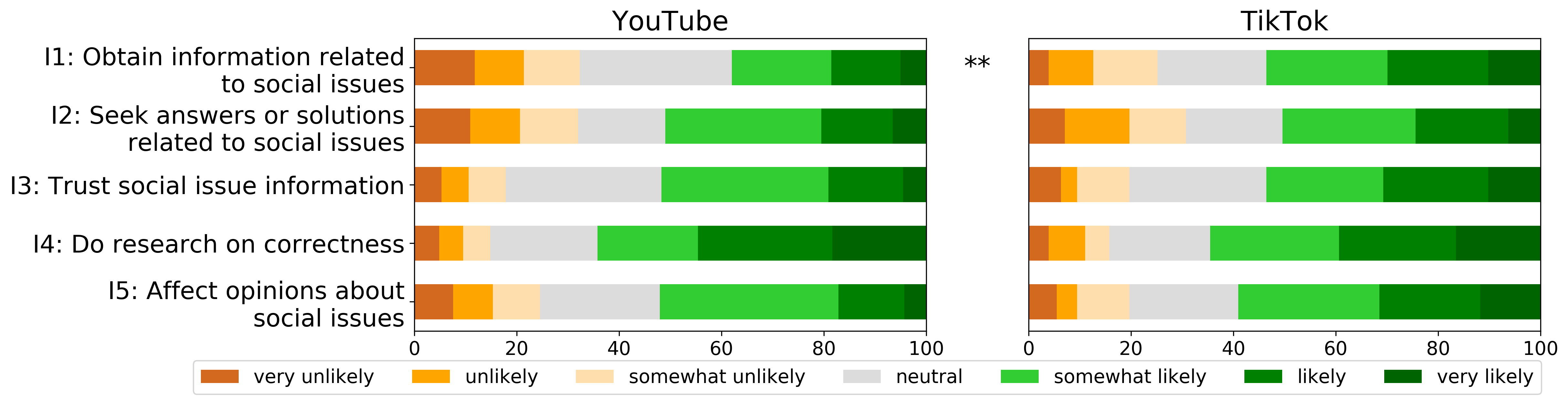}
        }\\
        \scalebox{0.7}{
            \begin{tabular}{p{2cm}p{.3\linewidth}p{.3\linewidth}}
            \toprule
                &\multicolumn{2}{l}{\textbf{I1: Obtain info related to social issue}}\\
                & \multicolumn{2}{l}{$F=6.39$,  $r^2=0.16$,  $p<0.0001$} \\
                \midrule
                factor & $F$ ratio & $p$ \\
                Platform & 7.36 & 0.0068 \\
                Age & - & - \\
                Gender & - & - \\
                Ethnicity & 5.80 & $<$0.0001 \\
                Frequency & 11.01 & $<$0.0001 \\
            \bottomrule
            \end{tabular}
        }
    \end{tabular}
    \caption{The distribution and statistic results of ratings for questions about information seeking with social issue videos (I1 - I5). Only statistics with a significant p-value are presented in the table. $p$* $<$ 0.05, $p*<0.05$, $p**0.01$, $p***<0.001$. Adjusted $\alpha=0.01$.}
    \label{fig:info_seeking}
\end{figure}

I1 to I5 examine how YouTube and TikTok users seek information related to social issues and whether users trust the social issue videos (SIVs) they have watched (Figure \ref{fig:info_seeking}). I1 asks to what degree participants obtain social issue information on VSPs (1-very unlikely, 7-very likely). YouTube users rated slightly negative on average ($MEAN_{YouTbe}=3.96$, $SD=1.66$), while TikTok users rated this question positive ($MEAN_{TikTok}=4.52$, $SD=1.59$). I2 asks whether participants actively seek answers or solutions to social issues by watching online videos (1-very unlikely, 7-very likely). Users of both platforms gave positive ratings ($MEAN_{YouTube}=4.14$, $SD=1.71$, $MEAN_{TikTok}=4.24$, $SD=1.66$). The LSR model suggests the platform is a significant predictor (Figure \ref{fig:info_seeking}), indicating that TikTok users are more potential to watch SIVs to obtain social issue information than YouTube users (posthoc $p_{Welch}=0.0004$). I3 inquiries whether participants trusted the information, and I4 asks whether participants researched the social issues with other sources after watching (1-very unlikely, 7-very likely). Users of both platforms rated that they tended to trust the videos ($MEAN_{YouTube}=4.42$, $SD=1.38$, $MEAN_{TikTok}=4.59$, $SD=1.56$). However, they would still do research with other sources to verify the information correctness ($MEAN_{YouTube}=4.98$, $SD=1.62$, $MEAN_{YouTube}=4.90$, $SD=1.60$). There are no significant differences in the ratings of I3 and I4 between the two groups. I5 further asks if watching SIVs affects their opinions on the subject (1-very unlikely, 7-very likely). Participants gave a moderately positive rating of 4.26 for YouTube ($SD=1.51$) and 4.68 for TikTok ($SD=1.56$, no significant difference between the two groups). 
\par
These results suggest that although YouTube and TikTok users are motivated by seeking information on VSPs, obtaining social issue information (I1) and seeking solutions related to social issues (I2) are not highly rated motives. Interestingly, although YouTube users gave higher ratings for information-related gratifications, TikTok users are more likely to obtain social issue information from SIVs. This could result from TikTok users passively watching videos through the ``\textit{For You}'' feature. Users of both platforms tend to trust the social issue information in the videos (I3), and the videos could affect their opinions (I4). However, users would still research to verify the information in SIVs (I5). 
\par

\begin{figure}[!h]
    \centering
    \scalebox{0.95}{
            \includegraphics[width=\linewidth]{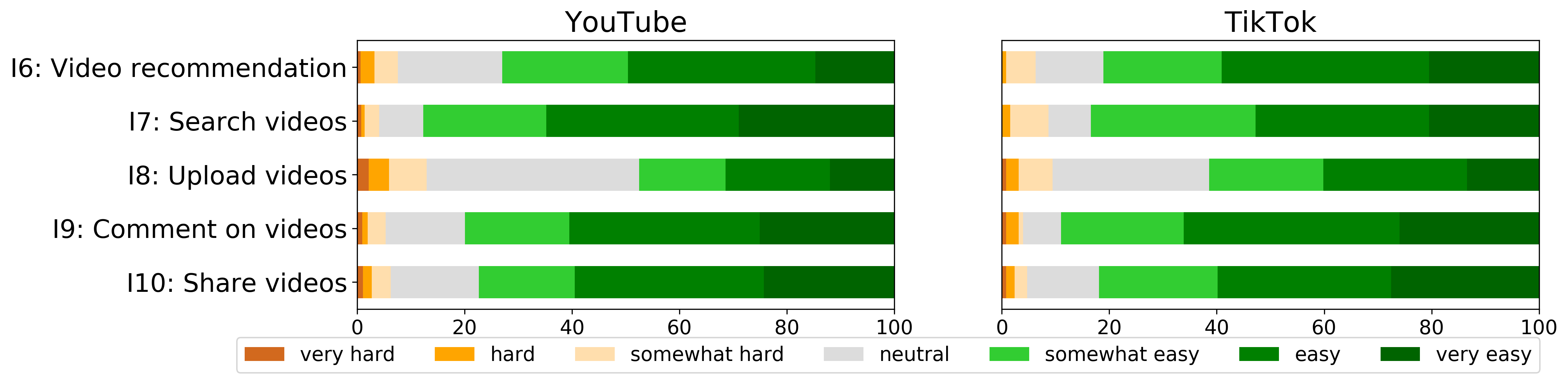}
        }\\
        \scalebox{0.7}{
            \begin{tabular}{p{2cm}p{0.12\linewidth}p{0.12\linewidth}|p{0.12\linewidth}p{0.12\linewidth}|p{0.12\linewidth}p{0.12\linewidth}|p{0.12\linewidth}p{0.12\linewidth}|p{0.12\linewidth}p{0.12\linewidth}}
            \toprule
                &\multicolumn{2}{l}{\textbf{I6: Recommendation}} &\multicolumn{2}{l}{\textbf{I7: Search}} &\multicolumn{2}{l}{\textbf{I8: Upload}} &\multicolumn{2}{l}{\textbf{I9: Comment}} &\multicolumn{2}{l}{\textbf{I10: Share}}\\
                \midrule
                Platform & $MEAN$ & $SD$ & $MEAN$ & $SD$ & $MEAN$ & $SD$ & $MEAN$ & $SD$ & $MEAN$ & $SD$\\
                YouTube & 5.26 & 1.25 & 5.75 & 1.45 & 4.70 & 1.39 & 5.57 & 1.25 & 5.51 & 1.30 \\
                TikTok & 5.54 & 1.16 & 5.46 & 1.20 & 5.02 & 1.30 & 5.73 & 1.16 & 5.61 & 1.25\\
            \bottomrule
            \end{tabular}
        }
    \caption{The distribution and statistic results of ratings for questions about the convenient utility of social issue videos (I6 - I10).}
    \label{fig:conv_util}
\end{figure}

I6 to I10 exam the convenient utility theme (Figure \ref{fig:conv_util}). We asked how efficiently the common features on both platforms affect the experiences of interacting with SIVs (1-very hard, 7-very easy). Except for uploading videos on YouTube, all other categories received positive ratings above 5. No significant difference exists between the two groups in any convenient utility questions.
\par

\begin{figure}[!h]
    \centering
    \begin{tabular}{c}
        \scalebox{0.95}{
            \includegraphics[width=\linewidth]{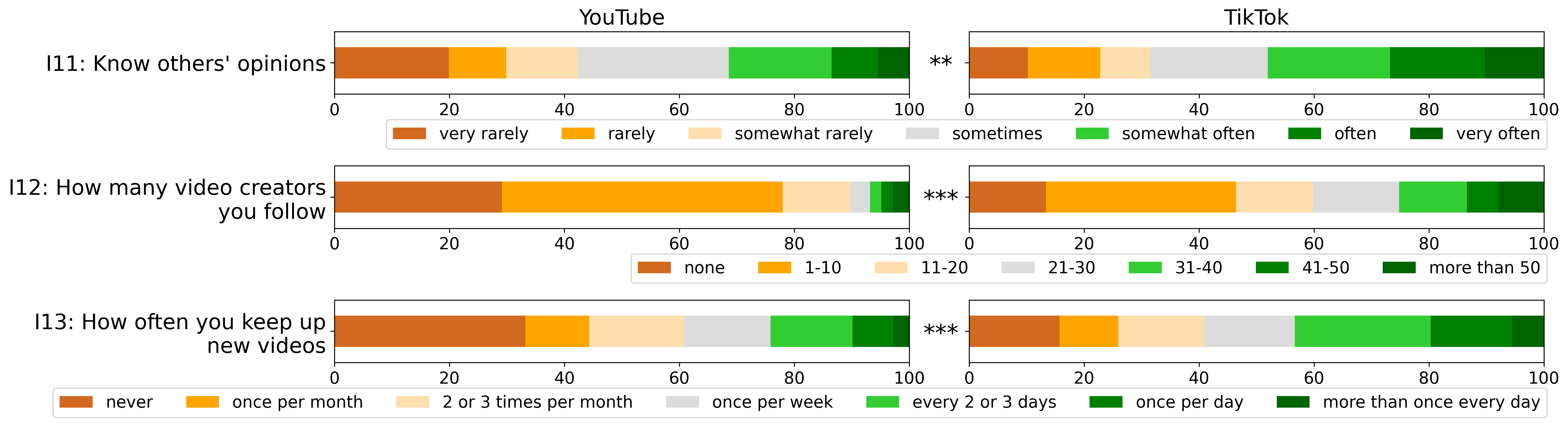}
        }\\
        \scalebox{0.7}{
            \begin{tabular}{p{2cm}p{0.15\linewidth}p{0.15\linewidth}|p{0.15\linewidth}p{0.15\linewidth}|p{0.15\linewidth}p{0.15\linewidth}}
            \toprule
                &\multicolumn{2}{l}{\textbf{I11: Know others' opinions}}&\multicolumn{2}{l}{\textbf{I12: Follow creators}}&\multicolumn{2}{l}{\textbf{I13: Keep up new videos}}\\
                & \multicolumn{2}{l}{$F=9.17$,  $r^2=0.21$,  $p<0.0001$}& \multicolumn{2}{l}{$F=9.76$,  $r^2=0.22$,  $p<0.0001$} \\               
                \midrule
                factor & $F$ ratio & $p$ & $F$ ratio & $p$ & $F$ ratio & $p$ \\
                Platform & 6.69 & $<$0.0001 & 55.47 & $<$0.0001 & 22.44 & $<$0.0001\\
                Age & - & - & 2.51 & 0.0208 & - & - \\
                Gender & - & -& - & - & - & -\\
                Ethnicity & 3.98 & 0.0003 & 2.48 & 0.0158 & 2.92 & 0.0051 \\
                Frequency & 16.13 & $<$0.0001 & 55.47 & $<$0.0001 & 22.44 & $<$0.0001 \\
            \bottomrule
            \end{tabular}
        }
    \end{tabular}
    \caption{The distribution and statistic results of ratings for questions about obtaining knowledge about others with social issue videos (I11 - I13). Only statistics with a significant p-value are presented in the table. $p*<0.05$, $p**0.01$, $p***<0.001$. Adjusted $\alpha=0.0167$.}
    \label{fig:info_knowledge_others}
\end{figure}

I11 to I13 examine to what degree VSPs are used for obtaining knowledge about others' opinions on social issues (Figure \ref{fig:info_knowledge_others}). I11 asks how often users use VSPs to know others' opinions on social issues (1-very rarely, 7-very often). YouTube users had an average of 3.58 ($SD=1.77$), while TikTok users had a positive rating of 4.2 ($SD=1.81$). The LSR model that predicts I11 shows a significant difference between the two platforms (Figure \ref{fig:info_knowledge_others}); TikTok users are more likely to use the platform to know others' opinions than YouTube users (posthoc $p_{Welch}=0.0005$). I12 asks how many video creators of SIVs the participants have followed (1-none, 7-more than 50), and I13 asks how often they keep up with creators' latest videos related to social issues (1-never, 7-more than once every day). YouTube users rated 2.18 on average ($SD=1.32$, 1-20 videos). TikTok users' average was 3.27 ($SD=1.79$, 11-30 videos). The LSR model suggests TikTok users follow more social issue creators than YouTube users (posthoc $p_{Welch}<0.0001$, Figure \ref{fig:info_knowledge_others}). Besides following, TikTok users are also more active in keeping up with new videos. The average rating of I13 of YouTube users was 2.98 ($SD=1.80$), between once per month and 2 or 3 times per month. The rating of TikTok users was 3.86 ($SD=1.81$), between 2 or 3 times per month and once per week. The LSR model also shows a significant difference between the two platforms (Figure \ref{fig:info_knowledge_others}) -- TikTok users are more likely to keep up the SIVs than YouTube users (posthoc $p_{Welch}<0.0001$). YouTube and TikTok are not often used to know and follow up on others' opinions. But TikTok users use the platform to know others' opinions and follow more creators than YouTube users. 
\par

\subsubsection{Entertainment Gratification and Social Issue Videos}
\begin{figure}[!h]
    \centering
    \begin{tabular}{c}
        \scalebox{0.95}{
            \includegraphics[width=\linewidth]{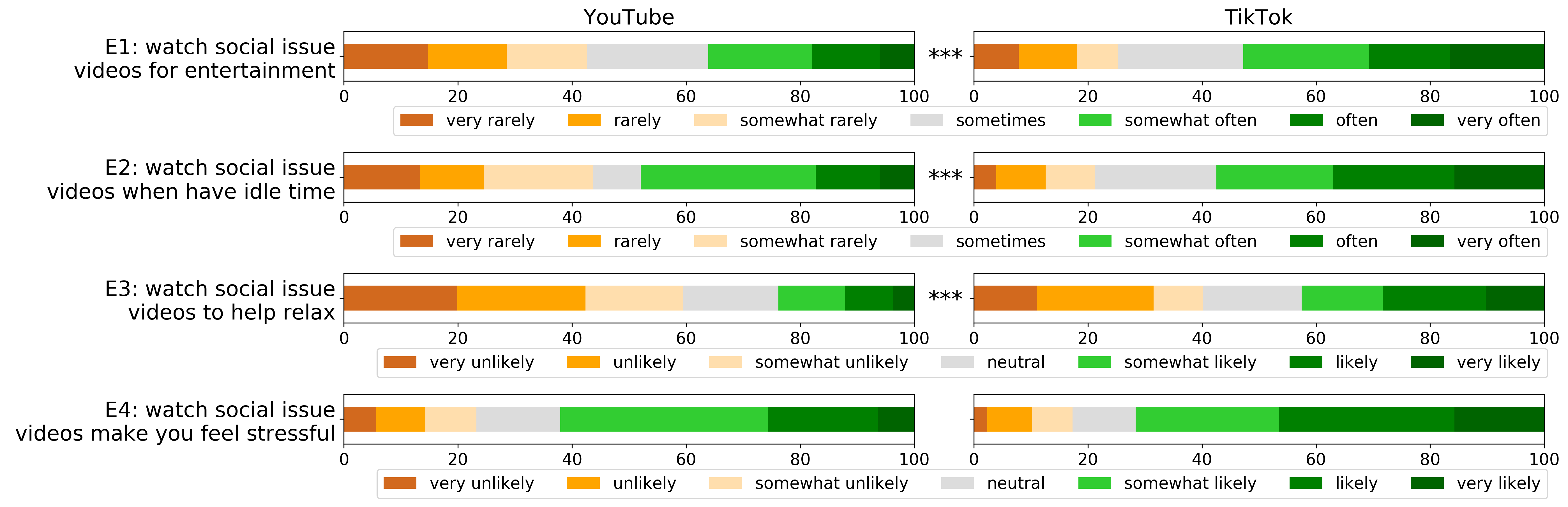}
        }\\
        \scalebox{0.7}{
            \begin{tabular}{p{1.5cm}p{0.14\linewidth}p{0.14\linewidth}|p{0.14\linewidth}p{0.14\linewidth}|p{0.14\linewidth}p{0.14\linewidth}}
            \toprule
                &\multicolumn{2}{l}{\textbf{E1: For entertainment}}&\multicolumn{2}{l}{\textbf{E2: Pass time}}&\multicolumn{2}{l}{\textbf{E3: Help relax}}\\
                & \multicolumn{2}{l}{$F=5.84$,  $r^2=0.15$,  $p<0.0001$}& \multicolumn{2}{l}{$F=6.93$,  $r^2=0.17$,  $p<0.0001$}& \multicolumn{2}{l}{$F=5.38$,  $r^2=0.14$,  $p<0.0001$}\\
                \midrule
                factor & $F$ ratio & $p$ & $F$ ratio & $p$ \\
                Platform & 12.58 & 0.0004 & 15.92 & <0.0001 & 18.38 & <0.0001\\
                Age & - & - & - & - & - & -\\
                Gender & - & - & - & - & - & -\\
                Ethnicity & 3.32 & 0.0017 & 3.30 & 0.0018 & - & -\\
                Frequency & 11.87 & <0.0001 & 15.92 & <0.0001 & 11.00 & <0.0001\\
            \bottomrule
            \end{tabular}
        }
    \end{tabular}
    \caption{The distribution and statistic results of ratings for entertainment with social issue videos (E1 - E4). Only statistics with a significant p-value are presented in the table. $p*<0.05$, $p**0.01$, $p***<0.001$. Adjusted $\alpha=0.0125$.}
    \label{fig:relaxation}
\end{figure}

The entertainment theme in the uses and gratifications theory examines whether users watch SIVs for entertainment, to pass the time, and feel relaxed after watching SIVs (Figure \ref{fig:relaxation}). E1 asks how often users watch SIVs for entertainment (1-very rarely, 7-very often). YouTube users gave a negative average rating ($MEAN_{YouTube}=3.74$, $SD=1.78$), while TikTok users gave a positive average rating ($MEAN_{TikTok}=4.49$, $SD=1.80$). Similarly, E2 asks how often users watch SIVs when idle (1-very rarely, 7-very often). This question had a negative average from YouTube users ($MEAN_{YouTube}=3.90$, $SD=1.77$) and a positive average from TikTok users ($MEAN_{TikTok}=4.72$, $SD=1.66$). The LSR models that predict E1 and E2 suggest platform is a significant predictor, indicating TikTok users are more likely to watch SIVs for entertainment and pass the idle time (both posthoc $p_{Welch}<0.0001$, Figure \ref{fig:relaxation}). E3 and E4 ask whether SIVs on VSP let users feel relaxed or stressed (1-very unlikely, 7-very likely). Both YouTube and TikTok users rated they felt less relaxed ($MEAN_{YouTube}=3.18$, $SD=1.72$, $MEAN_{TikTok}=3.98$, $SD=1.93$) and more stressed ($MEAN_{YouTube}=4.51$, $SD=1.55$, $MEAN_{TikTok}=5.04$, $SD=1.56$). The LSR model that predicts the E3 shows the platform factor is a significant predictor. Compared to TikTok users, YouTube users are less likely to feel relaxed watching SIVs (posthoc $p_{Welch}<0.0001$). YouTube users do not intend to watch SIVs for entertainment, killing time, or relaxation. On the contrary, TikTok users are more likely to watch SIVs for entertainment and pass the time than YouTube users. However, users of both platforms feel the SIVs are less relaxing but more stressful. 
\par

\subsubsection{Social Gratification and Social Issue Videos}
\begin{figure}
    \centering
    \begin{tabular}{c}
        \scalebox{0.95}{
            \includegraphics[width=\linewidth]{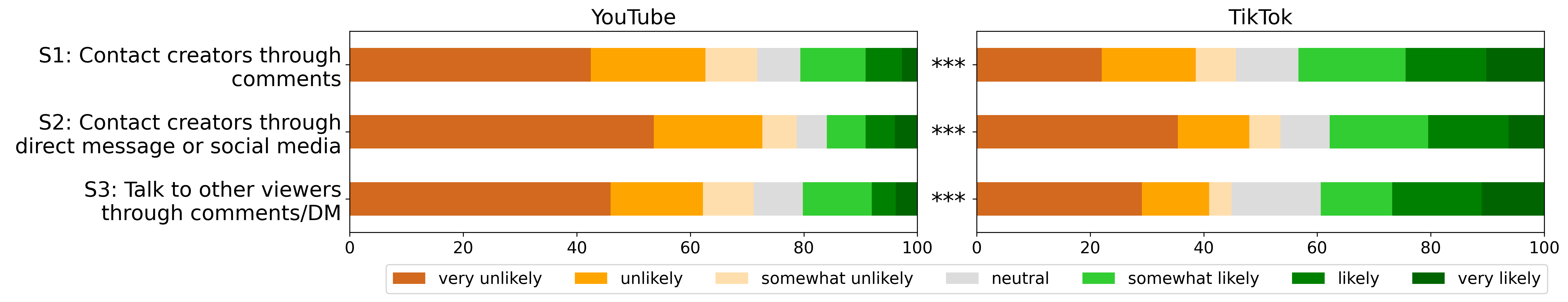}
        }\\
        \scalebox{0.7}{
            \begin{tabular}{p{2cm}p{0.15\linewidth}p{0.15\linewidth}|p{0.15\linewidth}p{0.15\linewidth}|p{0.15\linewidth}p{0.15\linewidth}}
            \toprule
                &\multicolumn{2}{l}{\textbf{S1: Contact through comments}}&\multicolumn{2}{l}{\textbf{S2: Contact through DM}}&\multicolumn{2}{l}{\textbf{S3: Talk to other viewers}}\\
                & \multicolumn{2}{l}{$F=8.25$,  $r^2=0.19$,  $p<0.0001$} & \multicolumn{2}{l}{$F=7.69$,  $r^2=0.18$,  $p<0.0001$} & \multicolumn{2}{l}{$F=8.41$,  $r^2=0.20$,  $p<0.0001$} \\
                \midrule
                factor & $F$ ratio & $p$ & $F$ ratio & $p$ & $F$ ratio & $p$\\
                Platform & 34.54 & $<$0.0001 & 27.55 & $<$0.0001 & 28.94 & $<$0.0001 \\
                Age & - & - & - & - & - & - \\
                Gender & - & - & - & - & - & -\\
                Ethnicity & 4.08 & 0.0002 & 2.89 & 0.0055 & 6.53 & $<$0.0001 \\
                Frequency & 14.83 & $<$0.0001 & 16.09 & $<$0.0001 &  14.48 & $<$0.0001 \\
            \bottomrule
            \end{tabular}
        }
    \end{tabular}
    \caption{The distribution and statistic results of ratings for social interaction with social issue videos (S1 - S3). Only statistics with a significant p-value are presented in the table. $p*<0.05$, $p**0.01$, $p***<0.001$. Adjusted $\alpha=0.0071$.}
    \label{fig:social_1}
\end{figure}

S1 to S3 examine the social interactions around SIVs (Figure \ref{fig:social_1}). S1 and S2 are questions regarding how likely users are to contact the creator through comments or other methods such as a direct message or other social media (1-very unlikely, 7-very likely). Both YouTube and TikTok users rated that they rarely contacted the creators via comments ($MEAN_{YouTube}=2.56$ $SD=1.80$, $MEAN_{TikTok}=3.72$, $SD=2.09$), direct messages, and social media ($MEAN_{YouTube}=2.24$ $SD=1.78$, $MEAN_{TikTok}=3.28$, $SD=2.14$). S3 asks if users talk to other viewers (1-very unlikely, 7-very likely). YouTube users rated 2.53 on average ($SD=1.83$), and TikTok users rated 3.62 on average ($SD=2.19$). The LSR models that predict S1, S2, and S3 show that platform is a significant predictor (all posthoc $p_{Welch}<0.0001$, Figure \ref{fig:social_1}). VSP users tend not to communicate with the creators or other viewers when watching SIVs. But TikTok users are more active in social interactions than YouTube users. 
\par

\begin{figure}
    \centering
    \begin{tabular}{c}
        \scalebox{0.95}{
            \includegraphics[width=\linewidth]{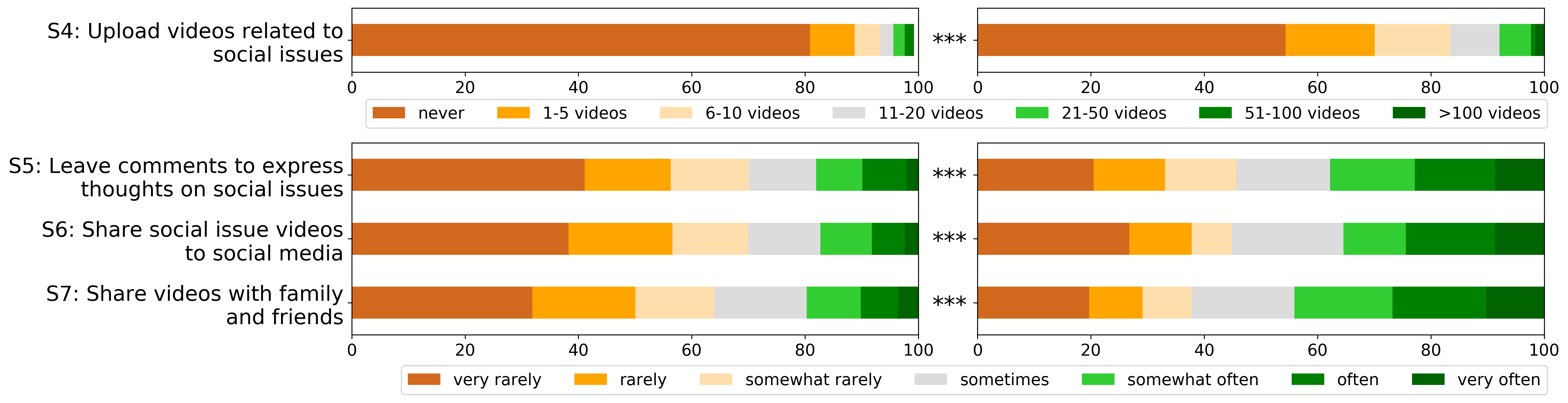}
        }\\
        \scalebox{0.7}{
            \begin{tabular}{p{2cm}p{0.15\linewidth}p{0.15\linewidth}|p{0.15\linewidth}p{0.15\linewidth}|p{0.15\linewidth}p{0.15\linewidth}|p{0.15\linewidth}p{0.15\linewidth}}
            \toprule
                & \multicolumn{2}{l}{\textbf{S4: Upload videos}} & \multicolumn{2}{l}{\textbf{S5: Express thoughts}} & \multicolumn{2}{l}{\textbf{S6: Share on social media}} & \multicolumn{2}{l}{\textbf{S7: Share with family}}\\
                & \multicolumn{2}{l}{$F=7.39$,  $r^2=0.18$,  $p<0.0001$} & \multicolumn{2}{l}{$F=8.80$,  $r^2=0.20$,  $p<0.0001$} & \multicolumn{2}{l}{$F=7.62$,  $r^2=0.18$,  $p<0.0001$} & \multicolumn{2}{l}{$F=7.74$,  $r^2=0.18$,  $p<0.0001$}\\
                \midrule
                factor & $F$ ratio & $p$ & $F$ ratio & $p$ & $F$ ratio & $p$ & $F$ ratio & $p$\\
                Platform & 28.10 & $<$0.0001 & 30.77 & $<$0.0001 & 21.65 & $<$0.0001 & 25.75 & $<$0.0001\\
                Age & 2.98 & 0.0070 & - & - & - & - & - & -\\
                Gender & - & - & - & - & - & - & - & -\\
                Ethnicity & - & - & 5.79 & $<$0.0001 & 4.46 & $<$0.0001 & 3.81 & 0.0004\\
                Frequency & 15.51 & $<$0.0001 & 15.90 & $<$0.0001 & 14.23 & $<$0.0001 & 13.75 & $<$0.0001\\
            \bottomrule
            \end{tabular}
        }
    \end{tabular}
    \caption{The distribution and statistic results of ratings for expressing and sharing opinions and communicatory utility with social issue videos (S4 - S7). Only statistics with a significant p-value are presented in the table. $p*<0.05$, $p**<0.01$, $p***<0.001$. Adjusted $\alpha=0.0071$.}
    \label{fig:social_2}
\end{figure}

S4 and S5 explore how VSP users express and share their opinions (Figure \ref{fig:social_2}). S4 asks whether the participants have uploaded videos related to social issues (1-never, 7-more than 100 videos). The average ratings of YouTube and TikTok users were 1.41 ($SD=1.03$) and 2.04 ($SD=1.33$). 533 out of 659 YouTube users (80.88\%) and 69 out of 127 TikTok users (54.33\%) have never shared any videos related to social issues. However, the LSR model that predicts S4 suggests TikTok users shared more videos on social issues than YouTube users (posthoc $p_{Welch}<0.0001$). S5 asks if participants express their thoughts on social issues through commenting on videos (1-very rarely, 7-very often). The result shows that YouTube users rated 2.63 ($SD=1.77$) and TikTok users rated 3.70 ($SD=1.98$), indicating users of both platforms are less often commenting on videos to express thoughts on social issues (posthoc $p_{Welch}<0.0001$). Alike sharing SIVs, the LSR model that predicts S5 suggests the platform is a significant predictor, with TikTok users being more likely to comment than YouTube users. Most VSP users do not share videos about social issues and do not actively leave comments on the videos. This indicates that VSPs are less used to discussing and debating social issues. But TikTok users are more likely to express opinions than YouTube users.
\par

S6 and S7 probe the communicatory utility of VSP videos for sharing information related to social issues (Figure \ref{fig:social_2}). S6 asks how often participants share SIVs on other social media (1-very rarely, 7-very often). YouTube users gave S6 an average rating of 2.63 ($SD=1.73$), and TikTok users gave it 3.59 ($SD=2.08$). The LSR model that predicts S6 shows that TikTok users share SIVs more often on other social media than YouTube users (posthoc $p_{Welch}<0.0001$). S7 asks how often participants share SIVs with family and friends (1-very rarely, 7-very often). YouTube users rated 2.88 ($SD=1.78$), and TikTok users rated 3.94 ($SD=2.01$). The LSR model that predicts S7 also indicates that TikTok users are more likely to share videos with family and friends than YouTube users (posthoc $p_{Welch}<0.0001$). These results imply that YouTube and TikTok users do not actively spread videos to other social media and people around them. But TikTok users are more likely to share SIVs than YouTube users.

\subsection{RQ3: How Gratifications Affect Watching Social Issue Videos}
\begin{figure}[!h]
    \centering
    \scalebox{0.8}{
            \includegraphics[width=\linewidth]{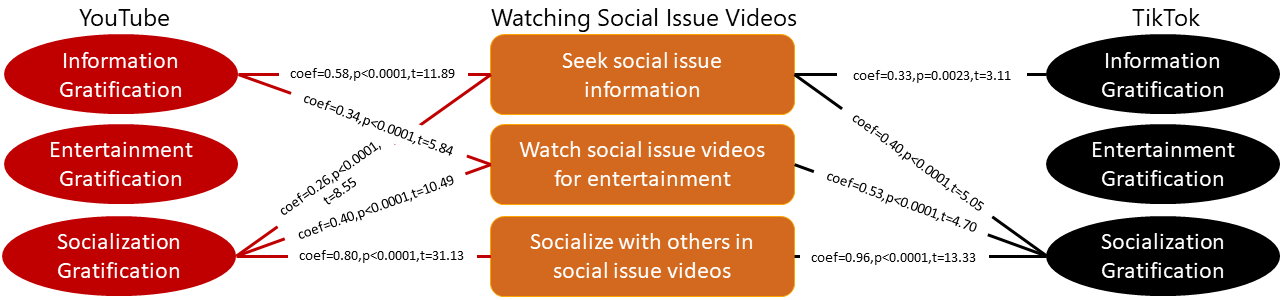}
        }
    \caption{The associations between the gratifications for video-sharing platforms and the ratings of uses of social issue videos.}
    \label{fig:RQ3}
\end{figure}

RQ3 examines if users' gratifications for VSPs affect their use of SIVs. LSR models use the aggregated scores for information, entertainment, and socialization gratifications to predict seeking information (average of I1 and I2), entertainment (E1), and social interaction (average of S1, S2, and S3) with SIVs. For seeking social issue information, the result suggests that information and socialization gratifications are significant predictors (see Figure \ref{fig:RQ3} for coefficients) for both YouTube users ($F=138.25$, $p<0.0001$, $R^2=0.39$) and TikTok users ($F=39.60$, $p<0.0001$, $R^2=0.49$). For YouTube users, the modal suggests that information and socialization gratifications significantly predict watching SIVs for entertainment ($F=93.36$, $p<0.0001$, $R^2=0.30$, Figure \ref{fig:RQ3}). While for TikTok, socialization is the sole significant predictor for entertainment with SIVs ($F=16.92$, $p<0.0001$, $R^2=0.29$, Figure \ref{fig:RQ3}). In the model that predicts socialization when watching SIVs, socialization gratification is the only significant predictor for both YouTube ($F=405.30$, $p<0.0001$, $R^2=0.65$) and TikTok ($F=131.15$, $p<0.0001$, $R^2=0.76$, Figure \ref{fig:RQ3}). 
\par
The results of RQ3 suggest that users who enjoy social interactions on VSPs are more likely to use SIVs for information-seeking, entertainment, and social interactions. VSP users who use YouTube or TikTok to obtain information tend to watch more SIVs. We also notice that entertainment gratification is not a significant predictor of entertainment with SIVs; there is no evidence that users who seek entertainment gratifications tend to watch SIVs for pleasure.

\subsection{RQ4: Perceptions of Watching Social Issue Videos}
\begin{figure}
    \centering
    \scalebox{0.7}{
            \includegraphics[width=\linewidth]{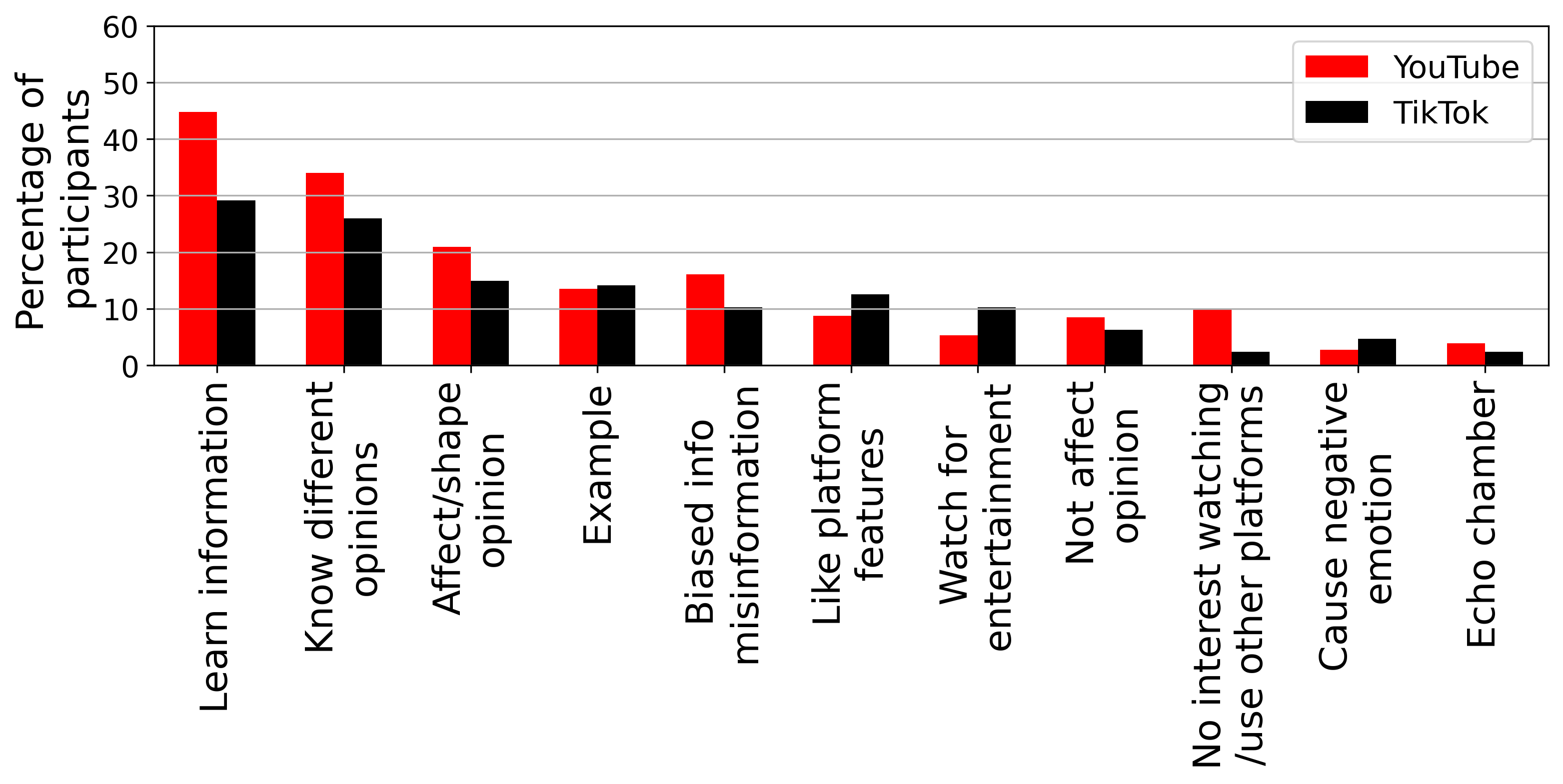}
        }
    \caption{Percentage of participants mention each of the RQ4 themes}
    \label{fig:reflection1}
\end{figure}

RQ4 is an open-ended question to probe how SIVs affect opinions. Participants were asked to write a short answer to reflect on how SIVs affect understanding, opinions, and actions on social issues. The thematic analysis identifies eleven themes from participants' feedback (Table \ref{fig:reflection1}). The top three themes are learning information, knowing different opinions, and affecting opinions. 
\par

\par
\begin{table*}[]
    \centering
    \scalebox{0.7}{
        \begin{tabular}{|p{0.22\linewidth}|p{0.32\linewidth}|p{0.8\linewidth}|}
        \hline
         RQ4 Theme & Definition & Examples \\
         \hline
         Learn information & Watch social issue videos to learn information and new knowledge & \textit{``If I hear an opinion or learn from youtube creators it is usually way easier to understand than just watching or reading the news.'' ``I have the freedom to seek out and learn more about different topics or even learn about current happenings that I otherwise wasn't even aware of.'}\\
         \hline
         Know different opinions & Watch social issue videos to know different views and perspectives & \textit{``They can provide you with insight on topics that would be deeper than other sources. You have a wide array of opinions to view.'' ``Mainly I can look at different videos and see two parts to the same social issue and make my own decision based on what I believe''}\\
         \hline
         Affect/shape opinion & Social issue videos affect opinions and understandings & \textit{``YouTube videos provide crucial insights that inform my decisions and opinions regarding social issues.'' ``I can see how it affects my thoughts and opinions on social issues.''}\\
         \hline
         Example & Give example video/channel/topic that the users watch & \textit{``Listening to PragerU videos has given me a better understanding about the history of the United States related to the good and the bad of the country's history.'' ``TikTok has brought to my attention to a county where the vast majority of residents don't have running water in the USA.''}\\
         \hline
         Biased information and misinformation & VSPs have biased information and mis/disinformation about social issues & \textit{``Some YouTube videos can be misleading.'' ``These issues affect all TikTokers, but they disproportionately impact kids who learn to define themselves and their self-worth based on such false realities.''}\\
         \hline
         Like platform features & Appreciate features of VSPs & \textit{``I like YouTube videos more than other apps. Because it's easily understandable to search and share options.'' ``Sometimes videos are more impactful than articles because visual images are more striking.''}\\
         \hline
         Watch for entertainment & Watch videos for fun and entertainment & \textit{``Just passed through the videos that are interesting. I have watched more viral videos.'' ``I feel like when I'm on TT I like to relax. I think that it is nice to see these once in a while.''}\\
         \hline
         Not affect opinion & Watching social issue videos do not affect opinions & \textit{``Youtube videos rarely if even change my mind on social issues.'' ``Most of the time, I don't take the TikTok video poster at their word and will do my own research''}\\
         \hline
         No interest in watching/using other platforms & No interest in watching social issues on VSPs or using other platforms & \textit{``I don't watch anything about social issues on Youtube.'' `` I actively avoid any content on the internet about things of this nature because I could not care any less about it.''}\\
         \hline
         Cause negative emotion & Watching social issue videos causes negative emotions or feelings & \textit{``I usually do not watch social issue videos because they cause me stress and are not fun.'' ``Depression, anxiety, anger, isolation, and extreme stress are all effects of cyberbullying.''}\\
         \hline
         Echo chamber & People only watch videos/channels that they agree with & \textit{``It serves only to reinforce my already existing opinions because I only watch things that agree with me.'' ``I think that the algorithm just keeps me in a bubble of my own opinions.''}\\
         \hline
        \end{tabular}
    }
    \caption{The themes and definitions identified from the thematic analysis of RQ4 responses}
    \label{tab:reflection1}
\end{table*}

295 (44.77\%) YouTube users and 37 (29.13\%) TikTok users mentioned using VSPs to obtain information about social issues. For example, one YouTube participant quoted, ``\textit{YouTube videos help to broaden my understanding of certain social issues. If other people's opinions differ from mine, I try to see their own points of view and understand why they think the way they do.}'' YouTube has 224 (33.99\%) users, and TikTok has 33 (25.98\%) users who noted that they watch SIVs to obtain diverse perspectives and opinions. For example, one YouTube user noted, ``\textit{YouTube videos are often a source of opinions and factual information discussed by various people. I tend to think the platform is just very good for finding out what others have to say about many different social issues.}'' Another TikTok user mentioned ``\textit{TikTok allows me to understand better what people think and what to happen in the future. I also learn a lot of information from people who have done the research. I think this is really important to keeping an open mind and forming your own opinion about a social issue.}'' Regarding whether watching SIVs affect opinions, 138 (20.94\%) YouTube users mentioned that online videos shaped or affected their opinions. In contrast, 56 YouTube (8.50\%) users noted the videos do not affect their opinions. Nineteen (14.96\%) TikTok users felt SIVs affected their views, while only eight (6.30\%) users noted the videos do not affect opinions. For example, one YouTube user noted ``\textit{A well-made video with a good articulation of facts will persuade me to think about the issues the way the creator intends.}'' A TikTok user quoted on how TikTok affects voting ``\textit{I remember watching videos and getting informed on how important this election is. I've seen people post videos of themselves voting; it really influenced me.}'' There were 58 (8.80\%) YouTube users and 16 (12.60\%) TikTok users who mentioned they liked the video forms or other platform features. As one YouTube user mentioned, ``\textit{YouTube can often provide a visual and auditory explanation for aspects of social issues that provide more details explanations than a simple article.}'' And another TikTok user quoted ``\textit{TikTok videos are very easy to watch due to their format, so I'm much more likely to pay attention and absorb information on social issues when learning about them via TikTok.}'' 
\par
Participants also commented on the negative perceptions and potential harms of watching SIVs. Among the participants, 106 (16.09\%) YouTube users and 13 (10.24\%) TikTok users noted biased information and misinformation about social issues on the platforms. For example, one YouTube user mentioned ``\textit{You will never see both sides of the argument unless your search is specific. A YouTube user must never assume that what is being claimed is true.}'' One TikTok user noted ``\textit{I really try to be aware, but I know a lot of the stuff I watch on there is just propaganda, and you have to do your own research.}'' There were 18 (2.73\%) YouTube users and 6 (4.72\%) TikTok users who expressed their negative emotions and feelings after watching SIVs. One YouTube user quoted ``\textit{Receiving so much bad information about certain issues makes me feel negative towards the future of society. It makes me feel upset at things that are happening and in the future.}'' Some participants (26 YouTube users and 3 TikTok users) commented about the echo chamber problems; they either noted that they only watch videos they agree with or that the platform feeds videos of similar opinions to the users. One YouTube user mentioned ``\textit{Echo chamber effect may cause others to believe their beliefs are a majority. It may even normalize extreme beliefs, resulting in a decline in critical analysis.}'' Another TikTok user noted ``\textit{TikTok strengthens my prior beliefs about certain issues and topics because I usually follow personalities on TikTok with the same opinion as mine.}''
\par

Participants watch SIVs because they can learn about social issues, obtain different opinions, and like watching videos. Participants also noted potential issues and concerns with SIVs on VSPs, including biased information and misinformation, negative emotions and feelings, and the echo chambers on the platform. 

\section{Discussion}
Our findings suggest how users of YouTube and TikTok watch social issue videos. In this section, we discuss the implications stemming from our survey.

\subsection{Obtaining Social Issue Information on Video-Sharing Platforms}
Similar to the findings in \cite{KhanSocialMediaEngagement, BalakrishnanYouTube, BufYouTube}, our results suggest seeking information is one of the primary motivators for VSP users. Information use and the convenience of information seeking are stronger motives for YouTube users than TikTok users. But regarding SIVs, YouTube users do not actively use the platform to obtain information related to social issues. In contrast, TikTok users are more likely to watch social issue videos and know others' opinions regarding social issues. TikTok users also follow more SIV creators and are more actively keeping up with new SIVs. Information gratification is a significant positive predictor of watching SIVs. Therefore, users using YouTube or TikTok for information-seeking are more likely to consume SIVs. YouTube users were found moderately interested in political videos and trust the content \cite{ZimmermannYouTube}. In our survey, users of both platforms tend to trust the social issue information and believe the videos could affect opinions. But both participant groups noted it was essential to do their own research to verify the correctness of the video content. Reflecting on how SIVs affect thoughts, participants rated SIVs as likely to affect their opinions on social issues. In the open-ended question, around 40\% of YouTube users and 30\% of TikTok users noted they use VSPs to learn information and know different opinions and perspectives. But meanwhile, participants are also concerned that SIVs could provide biased information and cause negative emotions and feelings. A few participants also noted the echo chamber issues on the platform.
\par
Our results suggest that VSP users consume social issue information on VSPs and use the videos to shape social and political opinions. Prior work argued that the extreme content results from the demand of users \cite{MungerSupplyDemand}; however, we found that VSP users appreciate the diverse views available on the platforms. Therefore, the design of algorithms and video recommendations should identify and present a broad spectrum of opinions and perspectives to allow users to gain different viewpoints. Although VSP users tend to trust the information offered by the creators they choose to watch, VSP users feel it necessary to do their own research. Therefore, new features that support performing user-led verification of legitimacy and correctness are needed. Designers should consider incorporating fact-checking tools to facilitate viewers to identify biased information and misinformation \cite{FanPlatformGovernance, PatilCredibilityIndicator, WohnContentCuration}. Our findings also indicate that TikTok users are more likely to obtain information related to social issues. This result could be due to using the ``\textit{For You}'' feature and watching the algorithm-recommended videos. It is critical to examine how TikTok video delivery mechanisms cause viewers' encounters with social issue information. Since TikTok tend to follow more SIV creators' videos than YouTube users, algorithm designers need to assess the weight of videos of the same person in the recommendation, as it may expose TikTok users to the opinions of the same creators.

\subsection{Entertainment and Social Issue Videos}
Entertainment was found to be a primary motive for using both YouTube \cite{KlobasYouTube, KhanSocialMediaEngagement, HaqquTeenagersYouTube} and TikTok \cite{YangTikTokChina, BossenUGTikTok, AhlseTikTok}. Participants were attracted by the entertaining content and would use the platform to pass the time and relax. Regarding SIVs, YouTube users rated SIVs less used for entertainment, while TikTok users tend to watch SIVs for entertainment. Users of both platforms watch SIVs when they have idle time. We also find that users who use VSPs to satisfy entertainment gratifications do not tend to watch SIVs for entertainment. Users of both platforms do not find the SIVs relaxing and rated it possible to cause stressful feelings. Around 5\% of participants also noted the videos cause negative emotions and feelings in the open-ended question. 
\par
These results suggest two main implications. First, the delivery of SIVs, especially on YouTube, should not be to satisfy the entertainment gratifications. Video recommendations that bring SIVs to users who mostly watch entertainment videos could cause adverse effects. When planning video content, YouTube and TikTok creators should consider viewers' entertainment motives. If the channel is designed for recreation and leisure, discussing topics related to ongoing social events and problems may cause negative feelings in the viewers. Second, creators could make SIVs more entertaining but less serious to satisfy users' entertainment gratifications. Creators may commodify the social issue topics, which could mislead the users. Research suggests 70\% of the most popular independent news channels are oriented around a personality \cite{StockingPewResearch}. Influential creators such as PewDiePie involve social discourse in humorous speech and entertainment content \cite{HokkaRacismYouTube}. Studies found that VSP videos lack deeper discussions on social issues \cite{NiuTeamTrees, Shapiro2014}. One explanation is that video creators dedicate their content to fulfill entertainment gratifications and would not involve serious social and political discussion. Future research should examine how entertaining video content affects viewers' trust in the information and opinions on social issues.

\subsection{Social Interactions with Social Issue Videos}
Studies found social interaction motives are connected to commenting and uploading activities on VSPs \cite{KhanSocialMediaEngagement, Balakrishnan2008DoCollaboration}. TikTok users were motivated by social interaction and networking on the platform \cite{OmarTikTok, BossenUGTikTok, LuTikTok}. Our findings suggest that the participants' socialization is a less primary gratification. But compared to YouTube, TikTok users are significantly more motivated by social interactions, expressing opinions, and sharing information. The higher gratification for social interactions affects how YouTube and TikTok users consume SIVs. Users with more social interactions on VSPs are more likely to seek social issue information, watch SIVs for entertainment, and socialize with others. TikTok users are more likely to contact creators and other viewers through comments, direct messages, or other social media than YouTube users. These social interactions could cultivate parasocial relationships with the creators. TikTok users are more active in uploading videos and expressing thoughts about social issues than YouTube users. TikTok users are more likely to spread SIVs by sharing them on other social media or with family and friends. 
\par
Social computing researchers have used video comments to obtain topics and opinions of the public \cite{OttoniRightWingYouTubeChannels, LangeYouTubeRants}. However, our study suggests that most viewers do not tend to express opinions under the videos, which indicates that video comments may not present all viewers who have watched the video. For users who socialize, parasocial interactions with YouTubers and TikTokers induce viewers to offer instrumental, emotional, and financial support \cite{WohnParasocialInteraction, YangTikTokChina}. Our survey results suggest that the social gratification of VSPs leads to higher social interactions with SIVs. TikTok users are more gratified by socializing with the creators and other viewers. Short-video platforms like TikTok offer new communicatory affordances. However, more social interactions with TikTokers could lead viewers to be more influenced by SIVs. Considering TikTok is more popular among the young generation, future research should examine the social activities on TikTok and discern how they affect viewers' opinions. Sharing short-form SIVs can also spread information to other platforms and influence families and friends. Studying the cross-platform interactions between VSPs and other social media will be necessary to understand how TikTok videos spread social issue information.

\section{Conclusion and Future Work}
In this work, we present a survey of 659 YouTube and 127 TikTok users to understand how people consume social issue videos (SIVs) on video-sharing platforms (VSPs). Our analysis finds that VSPs users are more attracted by information and entertainment gratifications. YouTube users are motivated by information-seeking gratifications more, while TikTok users rated entertainment and social gratifications higher. Regarding SIVs, YouTube users moderately obtain information from SIVs, while TikTok users are more likely to watch SIVs. Users of both platforms feel online videos affect opinions, but researching correctness is needed. YouTube users watch SIVs to kill time more than for entertainment. TikTok users watch SIVs for entertainment more often than YouTube users. Users of both platforms do not find SIVs relaxing but feel they are stressful. Users of both platforms do not actively have social interactions with SIVs. But compared to YouTube users, TikTok users are more likely to follow SIV creators and keep up with new videos. TikTok users also performed more social interactions with SIVs, such as leaving comments and sharing TikTok videos with others. 

\par
This work lays the foundations for our future research in three directions. From the information-seeking perspective, short-form, mobile-focused platforms like TikTok attracted users to interact with SIVs more often. Researchers need a deeper delve into the algorithms and video rankings that lead to the interactions with SIVs and discern their effects. From an entertainment perspective, it will be interesting to investigate how creators embed social issue information in entertainment channels and how to detect misinformation presented in entertainment videos. Third, from the social perspective, future work should examine how creators leverage parasocial interactions with viewers to spread information and express thoughts. Since social gratification is associated with watching SIVs, studying new video forms, such as user-created news, vlogs, or social experiments, will be valuable for understanding how emerging videos affect people's opinions.

\bibliographystyle{ACM-Reference-Format}
\bibliography{references.bib}


\begin{thebibliography}{70}

%%% the Web address.

\ifx \showCODEN    \undefined \def \showCODEN     #1{\unskip}     \fi
\ifx \showDOI      \undefined \def \showDOI       #1{#1}\fi
\ifx \showISBNx    \undefined \def \showISBNx     #1{\unskip}     \fi
\ifx \showISBNxiii \undefined \def \showISBNxiii  #1{\unskip}     \fi
\ifx \showISSN     \undefined \def \showISSN      #1{\unskip}     \fi
\ifx \showLCCN     \undefined \def \showLCCN      #1{\unskip}     \fi
\ifx \shownote     \undefined \def \shownote      #1{#1}          \fi
\ifx \showarticletitle \undefined \def \showarticletitle #1{#1}   \fi
\ifx \showURL      \undefined \def \showURL       {\relax}        \fi
% invisible to TeX
\providecommand\bibfield[2]{#2}
\providecommand\bibinfo[2]{#2}
\providecommand\natexlab[1]{#1}
\providecommand\showeprint[2][]{arXiv:#2}

\bibitem[\protect\citeauthoryear{Ahlse, Nilsson, and Sandstr{\"{o}}m}{Ahlse
  et~al\mbox{.}}{2020}]%
        {AhlseTikTok}
\bibfield{author}{\bibinfo{person}{Johannes Ahlse}, \bibinfo{person}{Felix
  Nilsson}, {and} \bibinfo{person}{Nina Sandstr{\"{o}}m}.}
  \bibinfo{year}{2020}\natexlab{}.
\newblock \showarticletitle{{It's time to TikTok : Exploring Generation Z's
  motivations to participate in {\#}Challenges}}.
\newblock   \bibinfo{volume}{Independen} (\bibinfo{year}{2020}),
  \bibinfo{pages}{77}.
\newblock
\urldef\tempurl%
\url{http://urn.kb.se/resolve?urn=urn:nbn:se:hj:diva-48708}
\showURL{%
\tempurl}


\bibitem[\protect\citeauthoryear{Allgaier}{Allgaier}{2019}]%
        {AllgaierYouTubeClimateChange}
\bibfield{author}{\bibinfo{person}{Joachim Allgaier}.}
  \bibinfo{year}{2019}\natexlab{}.
\newblock \showarticletitle{{Science and Environmental Communication on
  YouTube: Strategically Distorted Communications in Online Videos on Climate
  Change and Climate Engineering}}.
\newblock \bibinfo{journal}{\emph{Frontiers in Communication}}
  \bibinfo{volume}{4} (\bibinfo{year}{2019}).
\newblock
\showISSN{2297-900X}
\urldef\tempurl%
\url{https://doi.org/10.3389/fcomm.2019.00036}
\showDOI{\tempurl}


\bibitem[\protect\citeauthoryear{Balakrishnan, Fussell, and
  Kiesler}{Balakrishnan et~al\mbox{.}}{2008}]%
        {Balakrishnan2008DoCollaboration}
\bibfield{author}{\bibinfo{person}{Aruna~D. Balakrishnan},
  \bibinfo{person}{Susan~R. Fussell}, {and} \bibinfo{person}{Sara Kiesler}.}
  \bibinfo{year}{2008}\natexlab{}.
\newblock \showarticletitle{{Do visualizations improve synchronous remote
  collaboration?}}. In \bibinfo{booktitle}{\emph{Proceeding of the twenty-sixth
  annual CHI conference on Human factors in computing systems - CHI '08}}.
  \bibinfo{publisher}{ACM}, \bibinfo{address}{Florence, Italy},
  \bibinfo{pages}{1227}.
\newblock
\showISBNx{9781605580111}
\urldef\tempurl%
\url{https://doi.org/10.1145/1357054.1357246}
\showDOI{\tempurl}


\bibitem[\protect\citeauthoryear{Balakrishnan and Griffiths}{Balakrishnan and
  Griffiths}{2017}]%
        {BalakrishnanYouTube}
\bibfield{author}{\bibinfo{person}{Janarthanan Balakrishnan} {and}
  \bibinfo{person}{Mark~D Griffiths}.} \bibinfo{year}{2017}\natexlab{}.
\newblock \showarticletitle{{Social media addiction: What is the role of
  content in YouTube?}}
\newblock \bibinfo{journal}{\emph{Journal of Behavioral Addictions}}
  \bibinfo{volume}{6}, \bibinfo{number}{3} (\bibinfo{year}{2017}),
  \bibinfo{pages}{364--377}.
\newblock
\urldef\tempurl%
\url{https://doi.org/10.1556/2006.6.2017.058}
\showDOI{\tempurl}


\bibitem[\protect\citeauthoryear{Bartolome and Niu}{Bartolome and Niu}{2023}]%
        {NiuVSPLR}
\bibfield{author}{\bibinfo{person}{Ava Bartolome} {and} \bibinfo{person}{Shuo
  Niu}.} \bibinfo{year}{2023}\natexlab{}.
\newblock \showarticletitle{{A Literature Review of Video-Sharing Platform
  Research in HCI}}. In \bibinfo{booktitle}{\emph{Proceedings of the 2023 CHI
  Conference on Human Factors in Computing Systems}}
  \emph{(\bibinfo{series}{CHI '23})}. \bibinfo{publisher}{Association for
  Computing Machinery}, \bibinfo{address}{New York, NY, USA}.
\newblock
\showISBNx{978145039421}
\urldef\tempurl%
\url{https://doi.org/10.1145/3544548.3581107}
\showDOI{\tempurl}


\bibitem[\protect\citeauthoryear{Basch, Basch, Ruggles, and Hammond}{Basch
  et~al\mbox{.}}{2015}]%
        {BaschYouTubeEbola}
\bibfield{author}{\bibinfo{person}{Corey~H Basch}, \bibinfo{person}{Charles~E
  Basch}, \bibinfo{person}{Kelly~V Ruggles}, {and} \bibinfo{person}{Rodney
  Hammond}.} \bibinfo{year}{2015}\natexlab{}.
\newblock \showarticletitle{{Coverage of the Ebola virus disease epidemic on
  YouTube}}.
\newblock \bibinfo{journal}{\emph{Disaster medicine and public health
  preparedness}} \bibinfo{volume}{9}, \bibinfo{number}{5}
  (\bibinfo{year}{2015}), \bibinfo{pages}{531--535}.
\newblock
\showISSN{1935-7893}
\urldef\tempurl%
\url{https://doi.org/10.1017/dmp.2015.77}
\showDOI{\tempurl}


\bibitem[\protect\citeauthoryear{Basch, Meleo-Erwin, Fera, Jaime, and
  Basch}{Basch et~al\mbox{.}}{2021}]%
        {BaschCOVIDTikTok}
\bibfield{author}{\bibinfo{person}{Corey~H Basch}, \bibinfo{person}{Zoe
  Meleo-Erwin}, \bibinfo{person}{Joseph Fera}, \bibinfo{person}{Christie
  Jaime}, {and} \bibinfo{person}{Charles~E Basch}.}
  \bibinfo{year}{2021}\natexlab{}.
\newblock \showarticletitle{{A global pandemic in the time of viral memes:
  COVID-19 vaccine misinformation and disinformation on TikTok}}.
\newblock \bibinfo{journal}{\emph{Human Vaccines {\&} Immunotherapeutics}}
  \bibinfo{volume}{17}, \bibinfo{number}{8} (\bibinfo{date}{8}
  \bibinfo{year}{2021}), \bibinfo{pages}{2373--2377}.
\newblock
\showISSN{2164-5515}
\urldef\tempurl%
\url{https://doi.org/10.1080/21645515.2021.1894896}
\showDOI{\tempurl}


\bibitem[\protect\citeauthoryear{Basch, Yalamanchili, and Fera}{Basch
  et~al\mbox{.}}{2022}]%
        {BaschClimateChangeTikTok}
\bibfield{author}{\bibinfo{person}{Corey~H Basch}, \bibinfo{person}{Bhavya
  Yalamanchili}, {and} \bibinfo{person}{Joseph Fera}.}
  \bibinfo{year}{2022}\natexlab{}.
\newblock \showarticletitle{{{\#}Climate Change on TikTok: A Content Analysis
  of Videos}}.
\newblock \bibinfo{journal}{\emph{Journal of Community Health}}
  \bibinfo{volume}{47}, \bibinfo{number}{1} (\bibinfo{year}{2022}),
  \bibinfo{pages}{163--167}.
\newblock
\showISSN{1573-3610}
\urldef\tempurl%
\url{https://doi.org/10.1007/s10900-021-01031-x}
\showDOI{\tempurl}


\bibitem[\protect\citeauthoryear{Bora, Das, Barman, and Borah}{Bora
  et~al\mbox{.}}{2018}]%
        {BoraYouTubeZika}
\bibfield{author}{\bibinfo{person}{Kaustubh Bora}, \bibinfo{person}{Dulmoni
  Das}, \bibinfo{person}{Bhupen Barman}, {and} \bibinfo{person}{Probodh
  Borah}.} \bibinfo{year}{2018}\natexlab{}.
\newblock \showarticletitle{{Are internet videos useful sources of information
  during global public health emergencies? A case study of YouTube videos
  during the 2015–16 Zika virus pandemic}}.
\newblock \bibinfo{journal}{\emph{Pathogens and Global Health}}
  \bibinfo{volume}{112}, \bibinfo{number}{6} (\bibinfo{date}{8}
  \bibinfo{year}{2018}), \bibinfo{pages}{320--328}.
\newblock
\showISSN{2047-7724}
\urldef\tempurl%
\url{https://doi.org/10.1080/20477724.2018.1507784}
\showDOI{\tempurl}


\bibitem[\protect\citeauthoryear{Bowyer, Kahne, and Middaugh}{Bowyer
  et~al\mbox{.}}{2015}]%
        {BowyerYouTubePolitical}
\bibfield{author}{\bibinfo{person}{Benjamin~T Bowyer},
  \bibinfo{person}{Joseph~E Kahne}, {and} \bibinfo{person}{Ellen Middaugh}.}
  \bibinfo{year}{2015}\natexlab{}.
\newblock \showarticletitle{{Youth comprehension of political messages in
  YouTube videos}}.
\newblock \bibinfo{journal}{\emph{New Media {\&} Society}}
  \bibinfo{volume}{19}, \bibinfo{number}{4} (\bibinfo{date}{10}
  \bibinfo{year}{2015}), \bibinfo{pages}{522--541}.
\newblock
\showISSN{1461-4448}
\urldef\tempurl%
\url{https://doi.org/10.1177/1461444815611593}
\showDOI{\tempurl}


\bibitem[\protect\citeauthoryear{Braun and Clarke}{Braun and Clarke}{2012}]%
        {BraunThematicAnalysis}
\bibfield{author}{\bibinfo{person}{Virginia Braun} {and}
  \bibinfo{person}{Victoria Clarke}.} \bibinfo{year}{2012}\natexlab{}.
\newblock \bibinfo{title}{{Thematic analysis.}}
\newblock , \bibinfo{numpages}{57--71}~pages.
\newblock
\showISBNx{1-4338-1005-0 (Hardcover); 978-1-43381-005-3 (Hardcover)}
\urldef\tempurl%
\url{https://doi.org/10.1037/13620-004}
\showDOI{\tempurl}


\bibitem[\protect\citeauthoryear{Bucknell~Bossen, Kottasz, Bossen, Kottasz,
  Bucknell~Bossen, and Kottasz}{Bucknell~Bossen et~al\mbox{.}}{2020}]%
        {BossenUGTikTok}
\bibfield{author}{\bibinfo{person}{Christina Bucknell~Bossen},
  \bibinfo{person}{Rita Kottasz}, \bibinfo{person}{Christina~Bucknell Bossen},
  \bibinfo{person}{Rita Kottasz}, \bibinfo{person}{Christina Bucknell~Bossen},
  {and} \bibinfo{person}{Rita Kottasz}.} \bibinfo{year}{2020}\natexlab{}.
\newblock \showarticletitle{{Uses and gratifications sought by pre-adolescent
  and adolescent TikTok consumers}}.
\newblock \bibinfo{journal}{\emph{Young Consumers: Insight and Ideas for
  Responsible Marketers}} \bibinfo{volume}{21}, \bibinfo{number}{4}
  (\bibinfo{date}{1} \bibinfo{year}{2020}), \bibinfo{pages}{463--478}.
\newblock
\showISSN{1747-3616}
\urldef\tempurl%
\url{https://doi.org/doi:10.1108/YC-07-2020-1186}
\showDOI{\tempurl}


\bibitem[\protect\citeauthoryear{Buf and Ștef{\u{a}}niț{\u{a}}}{Buf and
  Ștef{\u{a}}niț{\u{a}}}{2020}]%
        {BufYouTube}
\bibfield{author}{\bibinfo{person}{Diana-Maria Buf} {and} \bibinfo{person}{Oana
  Ștef{\u{a}}niț{\u{a}}}.} \bibinfo{year}{2020}\natexlab{}.
\newblock \showarticletitle{{Uses and Gratifications of YouTube: A Comparative
  Analysis of Users and Content Creators}}.
\newblock \bibinfo{journal}{\emph{Romanian Journal of Communication and Public
  Relations; Vol 22 No 2 (2020): Qualitative Research in CommunicationDO -
  10.21018/rjcpr.2020.2.301}} (\bibinfo{date}{7} \bibinfo{year}{2020}).
\newblock
\urldef\tempurl%
\url{https://www.journalofcommunication.ro/index.php/journalofcommunication/article/view/301}
\showURL{%
\tempurl}


\bibitem[\protect\citeauthoryear{Buntain, Bonneau, Nagler, and Tucker}{Buntain
  et~al\mbox{.}}{2021}]%
        {BuntainYouTubeRecommendation}
\bibfield{author}{\bibinfo{person}{Cody Buntain}, \bibinfo{person}{Richard
  Bonneau}, \bibinfo{person}{Jonathan Nagler}, {and} \bibinfo{person}{Joshua~A
  Tucker}.} \bibinfo{year}{2021}\natexlab{}.
\newblock \showarticletitle{{YouTube Recommendations and Effects on Sharing
  Across Online Social Platforms}}.
\newblock \bibinfo{journal}{\emph{Proc. ACM Hum.-Comput. Interact.}}
  \bibinfo{volume}{5}, \bibinfo{number}{CSCW1} (\bibinfo{date}{4}
  \bibinfo{year}{2021}).
\newblock
\urldef\tempurl%
\url{https://doi.org/10.1145/3449085}
\showDOI{\tempurl}


\bibitem[\protect\citeauthoryear{Burgess and Green}{Burgess and Green}{2018}]%
        {Burgess2018YouTube:Culture}
\bibfield{author}{\bibinfo{person}{Jean Burgess} {and} \bibinfo{person}{Joshua
  Green}.} \bibinfo{year}{2018}\natexlab{}.
\newblock \bibinfo{booktitle}{\emph{{YouTube: Online video and participatory
  culture}}}.
\newblock \bibinfo{publisher}{John Wiley {\&} Sons},
  \bibinfo{address}{Cambridge, UK}.
\newblock
\showISBNx{1509533591}


\bibitem[\protect\citeauthoryear{Chau}{Chau}{2010}]%
        {YouTubeParticipatoryCulture}
\bibfield{author}{\bibinfo{person}{Clement Chau}.}
  \bibinfo{year}{2010}\natexlab{}.
\newblock \showarticletitle{{YouTube as a participatory culture}}.
\newblock \bibinfo{journal}{\emph{New Directions for Youth Development}}
  \bibinfo{volume}{2010}, \bibinfo{number}{128} (\bibinfo{year}{2010}),
  \bibinfo{pages}{65--74}.
\newblock
\urldef\tempurl%
\url{https://doi.org/10.1002/yd.376}
\showDOI{\tempurl}


\bibitem[\protect\citeauthoryear{Che, Ip, and Lin}{Che et~al\mbox{.}}{2015}]%
        {Che2015ACharacteristics}
\bibfield{author}{\bibinfo{person}{Xianhui Che}, \bibinfo{person}{Barry Ip},
  {and} \bibinfo{person}{Ling Lin}.} \bibinfo{year}{2015}\natexlab{}.
\newblock \showarticletitle{{A Survey of Current YouTube Video
  Characteristics}}.
\newblock \bibinfo{journal}{\emph{IEEE MultiMedia}} \bibinfo{volume}{22},
  \bibinfo{number}{2} (\bibinfo{year}{2015}), \bibinfo{pages}{56--63}.
\newblock
\showISSN{1941-0166 VO - 22}
\urldef\tempurl%
\url{https://doi.org/10.1109/MMUL.2015.34}
\showDOI{\tempurl}


\bibitem[\protect\citeauthoryear{Covington, Adams, and Sargin}{Covington
  et~al\mbox{.}}{2016}]%
        {CovingtonYouTubeRecommendations}
\bibfield{author}{\bibinfo{person}{Paul Covington}, \bibinfo{person}{Jay
  Adams}, {and} \bibinfo{person}{Emre Sargin}.}
  \bibinfo{year}{2016}\natexlab{}.
\newblock \showarticletitle{{Deep Neural Networks for YouTube
  Recommendations}}. In \bibinfo{booktitle}{\emph{Proceedings of the 10th ACM
  Conference on Recommender Systems}} \emph{(\bibinfo{series}{RecSys '16})}.
  \bibinfo{publisher}{Association for Computing Machinery},
  \bibinfo{address}{New York, NY, USA}, \bibinfo{pages}{191–198}.
\newblock
\showISBNx{9781450340359}
\urldef\tempurl%
\url{https://doi.org/10.1145/2959100.2959190}
\showDOI{\tempurl}


\bibitem[\protect\citeauthoryear{Eriksson~Krutr{\"{o}}k and
  {\AA}kerlund}{Eriksson~Krutr{\"{o}}k and {\AA}kerlund}{2022}]%
        {KrutrokRacisTikTok}
\bibfield{author}{\bibinfo{person}{Moa Eriksson~Krutr{\"{o}}k} {and}
  \bibinfo{person}{Mathilda {\AA}kerlund}.} \bibinfo{year}{2022}\natexlab{}.
\newblock \showarticletitle{{Through a white lens: Black victimhood,
  visibility, and whiteness in the Black Lives Matter movement on TikTok}}.
\newblock \bibinfo{journal}{\emph{Information, Communication {\&} Society}}
  (\bibinfo{date}{4} \bibinfo{year}{2022}), \bibinfo{pages}{1--19}.
\newblock
\showISSN{1369-118X}
\urldef\tempurl%
\url{https://doi.org/10.1080/1369118X.2022.2065211}
\showDOI{\tempurl}


\bibitem[\protect\citeauthoryear{Fan and Zhang}{Fan and Zhang}{2020}]%
        {FanPlatformGovernance}
\bibfield{author}{\bibinfo{person}{Jenny Fan} {and} \bibinfo{person}{Amy~X
  Zhang}.} \bibinfo{year}{2020}\natexlab{}.
\newblock \showarticletitle{{Digital Juries: A Civics-Oriented Approach to
  Platform Governance}}.
\newblock In \bibinfo{booktitle}{\emph{Proceedings of the 2020 CHI Conference
  on Human Factors in Computing Systems}}. \bibinfo{publisher}{Association for
  Computing Machinery}, \bibinfo{address}{New York, NY, USA},
  \bibinfo{pages}{1–14}.
\newblock
\showISBNx{9781450367080}
\urldef\tempurl%
\url{https://doi.org/10.1145/3313831.3376293}
\showURL{%
\tempurl}


\bibitem[\protect\citeauthoryear{{Google}}{{Google}}{2022}]%
        {YouTube5Year}
\bibfield{author}{\bibinfo{person}{{Google}}.} \bibinfo{year}{2022}\natexlab{}.
\newblock \bibinfo{title}{{History of Monetization at YouTube - YouTube5Year}}.
\newblock
\newblock
\urldef\tempurl%
\url{https://sites.google.com/a/pressatgoogle.com/youtube5year/home/history-of-monetization-at-youtube}
\showURL{%
\tempurl}


\bibitem[\protect\citeauthoryear{Guo and Harlow}{Guo and Harlow}{2014}]%
        {GuoRacismYouTube}
\bibfield{author}{\bibinfo{person}{Lei Guo} {and} \bibinfo{person}{Summer
  Harlow}.} \bibinfo{year}{2014}\natexlab{}.
\newblock \showarticletitle{{User-Generated Racism: An Analysis of Stereotypes
  of African Americans, Latinos, and Asians in YouTube Videos}}.
\newblock \bibinfo{journal}{\emph{Howard Journal of Communications}}
  \bibinfo{volume}{25}, \bibinfo{number}{3} (\bibinfo{date}{7}
  \bibinfo{year}{2014}), \bibinfo{pages}{281--302}.
\newblock
\showISSN{1064-6175}
\urldef\tempurl%
\url{https://doi.org/10.1080/10646175.2014.925413}
\showDOI{\tempurl}


\bibitem[\protect\citeauthoryear{Haqqu, Hastjarjo, and Slamet}{Haqqu
  et~al\mbox{.}}{2019}]%
        {HaqquTeenagersYouTube}
\bibfield{author}{\bibinfo{person}{Rizca Haqqu}, \bibinfo{person}{Sri
  Hastjarjo}, {and} \bibinfo{person}{Yulius Slamet}.}
  \bibinfo{year}{2019}\natexlab{}.
\newblock \showarticletitle{{Teenagers’ entertainment satisfaction in
  watching talk show program through YouTube}}.
\newblock \bibinfo{journal}{\emph{Jurnal The Messenger}} \bibinfo{volume}{11},
  \bibinfo{number}{1} (\bibinfo{year}{2019}), \bibinfo{pages}{38--45}.
\newblock
\showISSN{2527-2810}


\bibitem[\protect\citeauthoryear{Hautea, Parks, Takahashi, and Zeng}{Hautea
  et~al\mbox{.}}{2021}]%
        {HauteaTikTokClimateChange}
\bibfield{author}{\bibinfo{person}{Samantha Hautea}, \bibinfo{person}{Perry
  Parks}, \bibinfo{person}{Bruno Takahashi}, {and} \bibinfo{person}{Jing
  Zeng}.} \bibinfo{year}{2021}\natexlab{}.
\newblock \showarticletitle{{Showing They Care (Or Don’t): Affective Publics
  and Ambivalent Climate Activism on TikTok}}.
\newblock \bibinfo{journal}{\emph{Social Media + Society}} \bibinfo{volume}{7},
  \bibinfo{number}{2} (\bibinfo{date}{4} \bibinfo{year}{2021}),
  \bibinfo{pages}{20563051211012344}.
\newblock
\showISSN{2056-3051}
\urldef\tempurl%
\url{https://doi.org/10.1177/20563051211012344}
\showDOI{\tempurl}


\bibitem[\protect\citeauthoryear{He, He, Lu, and Li}{He et~al\mbox{.}}{2021}]%
        {HeChinaOnlineCrisisDanmaku}
\bibfield{author}{\bibinfo{person}{Changyang He}, \bibinfo{person}{Lu He},
  \bibinfo{person}{Tun Lu}, {and} \bibinfo{person}{Bo Li}.}
  \bibinfo{year}{2021}\natexlab{}.
\newblock \showarticletitle{{Beyond Entertainment: Unpacking Danmaku and
  Comments' Role of Information Sharing and Sentiment Expression in Online
  Crisis Videos}}.
\newblock \bibinfo{journal}{\emph{Proc. ACM Hum.-Comput. Interact.}}
  \bibinfo{volume}{5}, \bibinfo{number}{CSCW2} (\bibinfo{date}{10}
  \bibinfo{year}{2021}).
\newblock
\urldef\tempurl%
\url{https://doi.org/10.1145/3479555}
\showDOI{\tempurl}


\bibitem[\protect\citeauthoryear{Herrman}{Herrman}{2020}]%
        {HerrmanTikTok}
\bibfield{author}{\bibinfo{person}{John Herrman}.}
  \bibinfo{year}{2020}\natexlab{}.
\newblock \showarticletitle{{TikTok is shaping politics. But how}}.
\newblock \bibinfo{journal}{\emph{New York Times. https://www. nytimes.
  com/2020/06/28/style/tiktok-teen-politics-gen-z. html}}
  (\bibinfo{year}{2020}).
\newblock


\bibitem[\protect\citeauthoryear{Hokka}{Hokka}{2020}]%
        {HokkaRacismYouTube}
\bibfield{author}{\bibinfo{person}{Jenni Hokka}.}
  \bibinfo{year}{2020}\natexlab{}.
\newblock \showarticletitle{{PewDiePie, racism and Youtube’s neoliberalist
  interpretation of freedom of speech}}.
\newblock \bibinfo{journal}{\emph{Convergence}} \bibinfo{volume}{27},
  \bibinfo{number}{1} (\bibinfo{date}{7} \bibinfo{year}{2020}),
  \bibinfo{pages}{142--160}.
\newblock
\showISSN{1354-8565}
\urldef\tempurl%
\url{https://doi.org/10.1177/1354856520938602}
\showDOI{\tempurl}


\bibitem[\protect\citeauthoryear{Hou}{Hou}{2018}]%
        {Hou2018SocialYouTube}
\bibfield{author}{\bibinfo{person}{Mingyi Hou}.}
  \bibinfo{year}{2018}\natexlab{}.
\newblock \showarticletitle{{Social media celebrity and the
  institutionalization of YouTube}}.
\newblock \bibinfo{journal}{\emph{Convergence}} \bibinfo{volume}{25},
  \bibinfo{number}{3} (\bibinfo{date}{1} \bibinfo{year}{2018}),
  \bibinfo{pages}{534--553}.
\newblock
\showISSN{1354-8565}
\urldef\tempurl%
\url{https://doi.org/10.1177/1354856517750368}
\showDOI{\tempurl}


\bibitem[\protect\citeauthoryear{Hussein, Juneja, and Mitra}{Hussein
  et~al\mbox{.}}{2020}]%
        {MitraYouTubeMisinformation}
\bibfield{author}{\bibinfo{person}{Eslam Hussein}, \bibinfo{person}{Prerna
  Juneja}, {and} \bibinfo{person}{Tanushree Mitra}.}
  \bibinfo{year}{2020}\natexlab{}.
\newblock \showarticletitle{{Measuring Misinformation in Video Search
  Platforms: An Audit Study on YouTube}}.
\newblock \bibinfo{journal}{\emph{Proc. ACM Hum.-Comput. Interact.}}
  \bibinfo{volume}{4}, \bibinfo{number}{CSCW1} (\bibinfo{date}{5}
  \bibinfo{year}{2020}).
\newblock
\urldef\tempurl%
\url{https://doi.org/10.1145/3392854}
\showDOI{\tempurl}


\bibitem[\protect\citeauthoryear{Jhaver, Chen, Knauss, and Zhang}{Jhaver
  et~al\mbox{.}}{2022}]%
        {JhaverYouTubeCommentModeration}
\bibfield{author}{\bibinfo{person}{Shagun Jhaver}, \bibinfo{person}{Quan~Ze
  Chen}, \bibinfo{person}{Detlef Knauss}, {and} \bibinfo{person}{Amy~X Zhang}.}
  \bibinfo{year}{2022}\natexlab{}.
\newblock \showarticletitle{{Designing Word Filter Tools for Creator-Led
  Comment Moderation}}. In \bibinfo{booktitle}{\emph{CHI Conference on Human
  Factors in Computing Systems}} \emph{(\bibinfo{series}{CHI '22})}.
  \bibinfo{publisher}{Association for Computing Machinery},
  \bibinfo{address}{New York, NY, USA}.
\newblock
\showISBNx{9781450391573}
\urldef\tempurl%
\url{https://doi.org/10.1145/3491102.3517505}
\showDOI{\tempurl}


\bibitem[\protect\citeauthoryear{Katz, Blumler, and Gurevitch}{Katz
  et~al\mbox{.}}{1973}]%
        {KatzUGT}
\bibfield{author}{\bibinfo{person}{Elihu Katz}, \bibinfo{person}{Jay~G
  Blumler}, {and} \bibinfo{person}{Michael Gurevitch}.}
  \bibinfo{year}{1973}\natexlab{}.
\newblock \showarticletitle{{Uses and Gratifications Research}}.
\newblock \bibinfo{journal}{\emph{The Public Opinion Quarterly}}
  \bibinfo{volume}{37}, \bibinfo{number}{4} (\bibinfo{date}{5}
  \bibinfo{year}{1973}), \bibinfo{pages}{509--523}.
\newblock
\showISSN{0033362X, 15375331}
\urldef\tempurl%
\url{http://www.jstor.org/stable/2747854}
\showURL{%
\tempurl}


\bibitem[\protect\citeauthoryear{Khan}{Khan}{2017}]%
        {KhanSocialMediaEngagement}
\bibfield{author}{\bibinfo{person}{M~Laeeq Khan}.}
  \bibinfo{year}{2017}\natexlab{}.
\newblock \showarticletitle{{Social media engagement: What motivates user
  participation and consumption on YouTube?}}
\newblock \bibinfo{journal}{\emph{Computers in Human Behavior}}
  \bibinfo{volume}{66} (\bibinfo{year}{2017}), \bibinfo{pages}{236--247}.
\newblock
\showISSN{0747-5632}
\urldef\tempurl%
\url{https://doi.org/10.1016/j.chb.2016.09.024}
\showDOI{\tempurl}


\bibitem[\protect\citeauthoryear{Klobas, McGill, Moghavvemi, and
  Paramanathan}{Klobas et~al\mbox{.}}{2018}]%
        {KlobasYouTube}
\bibfield{author}{\bibinfo{person}{Jane~E Klobas}, \bibinfo{person}{Tanya~J
  McGill}, \bibinfo{person}{Sedigheh Moghavvemi}, {and}
  \bibinfo{person}{Tanousha Paramanathan}.} \bibinfo{year}{2018}\natexlab{}.
\newblock \showarticletitle{{Compulsive YouTube usage: A comparison of use
  motivation and personality effects}}.
\newblock \bibinfo{journal}{\emph{Computers in Human Behavior}}
  \bibinfo{volume}{87} (\bibinfo{year}{2018}), \bibinfo{pages}{129--139}.
\newblock
\showISSN{0747-5632}
\urldef\tempurl%
\url{https://doi.org/10.1016/j.chb.2018.05.038}
\showDOI{\tempurl}


\bibitem[\protect\citeauthoryear{Lange}{Lange}{2014}]%
        {LangeYouTubeRants}
\bibfield{author}{\bibinfo{person}{Patricia~G Lange}.}
  \bibinfo{year}{2014}\natexlab{}.
\newblock \showarticletitle{{Commenting on YouTube rants: Perceptions of
  inappropriateness or civic engagement?}}
\newblock \bibinfo{journal}{\emph{Journal of Pragmatics}}  \bibinfo{volume}{73}
  (\bibinfo{year}{2014}), \bibinfo{pages}{53--65}.
\newblock
\showISSN{0378-2166}
\urldef\tempurl%
\url{https://doi.org/10.1016/j.pragma.2014.07.004}
\showDOI{\tempurl}


\bibitem[\protect\citeauthoryear{Lee, Wu, Ertugrul, Lin, and Xie}{Lee
  et~al\mbox{.}}{2022}]%
        {LeeYouTubeControversialTopics}
\bibfield{author}{\bibinfo{person}{JooYoung Lee}, \bibinfo{person}{Siqi Wu},
  \bibinfo{person}{Ali~Mert Ertugrul}, \bibinfo{person}{Yu-Ru Lin}, {and}
  \bibinfo{person}{Lexing Xie}.} \bibinfo{year}{2022}\natexlab{}.
\newblock \showarticletitle{{Whose Advantage? Measuring Attention Dynamics
  across YouTube and Twitter on Controversial Topics}}.
\newblock \bibinfo{journal}{\emph{arXiv preprint arXiv:2204.00988}}
  (\bibinfo{year}{2022}).
\newblock


\bibitem[\protect\citeauthoryear{Lewis}{Lewis}{2019}]%
        {LewisYouTubeMicrocelebrity}
\bibfield{author}{\bibinfo{person}{Rebecca Lewis}.}
  \bibinfo{year}{2019}\natexlab{}.
\newblock \showarticletitle{{“This Is What the News Won’t Show You”:
  YouTube Creators and the Reactionary Politics of Micro-celebrity}}.
\newblock \bibinfo{journal}{\emph{Television {\&} New Media}}
  \bibinfo{volume}{21}, \bibinfo{number}{2} (\bibinfo{date}{10}
  \bibinfo{year}{2019}), \bibinfo{pages}{201--217}.
\newblock
\showISSN{1527-4764}
\urldef\tempurl%
\url{https://doi.org/10.1177/1527476419879919}
\showDOI{\tempurl}


\bibitem[\protect\citeauthoryear{Li, Pastukhova, Brandts-Longtin, Tan, and
  Kirchhof}{Li et~al\mbox{.}}{2022}]%
        {LiYouTubeCOVID}
\bibfield{author}{\bibinfo{person}{Heidi Oi-Yee Li}, \bibinfo{person}{Elena
  Pastukhova}, \bibinfo{person}{Olivier Brandts-Longtin},
  \bibinfo{person}{Marcus~G Tan}, {and} \bibinfo{person}{Mark~G Kirchhof}.}
  \bibinfo{year}{2022}\natexlab{}.
\newblock \showarticletitle{{YouTube as a source of misinformation on COVID-19
  vaccination: a systematic analysis}}.
\newblock \bibinfo{journal}{\emph{BMJ global health}} \bibinfo{volume}{7},
  \bibinfo{number}{3} (\bibinfo{year}{2022}).
\newblock
\showISSN{2059-7908}
\urldef\tempurl%
\url{https://doi.org/10.1136/bmjgh-2021-008334}
\showDOI{\tempurl}


\bibitem[\protect\citeauthoryear{Li, Guan, Hammond, and Berrey}{Li
  et~al\mbox{.}}{2021}]%
        {LiCOVIDTikTok}
\bibfield{author}{\bibinfo{person}{Yachao Li}, \bibinfo{person}{Mengfei Guan},
  \bibinfo{person}{Paige Hammond}, {and} \bibinfo{person}{Lane~E Berrey}.}
  \bibinfo{year}{2021}\natexlab{}.
\newblock \showarticletitle{{Communicating COVID-19 information on TikTok: a
  content analysis of TikTok videos from official accounts featured in the
  COVID-19 information hub}}.
\newblock \bibinfo{journal}{\emph{Health education research}}
  \bibinfo{volume}{36}, \bibinfo{number}{3} (\bibinfo{date}{7}
  \bibinfo{year}{2021}), \bibinfo{pages}{261--271}.
\newblock
\showISSN{1465-3648}
\urldef\tempurl%
\url{https://doi.org/10.1093/her/cyab010}
\showDOI{\tempurl}


\bibitem[\protect\citeauthoryear{Liu}{Liu}{2021}]%
        {LiuTikTokRacistTrend}
\bibfield{author}{\bibinfo{person}{Helen Liu}.}
  \bibinfo{year}{2021}\natexlab{}.
\newblock \showarticletitle{{Social Implications of Adolescents Engaging with
  Racist Trends on TikTok}}.
\newblock \bibinfo{journal}{\emph{The Media Education Research Journal}}
  (\bibinfo{date}{12} \bibinfo{year}{2021}).
\newblock
\urldef\tempurl%
\url{https://doi.org/10.5281/ZENODO.5763801}
\showDOI{\tempurl}


\bibitem[\protect\citeauthoryear{Lu and Lu}{Lu and Lu}{2019}]%
        {LuTikTok}
\bibfield{author}{\bibinfo{person}{Xing Lu} {and} \bibinfo{person}{Zhicong
  Lu}.} \bibinfo{year}{2019}\natexlab{}.
\newblock \showarticletitle{{Fifteen Seconds of Fame: A Qualitative Study of
  Douyin, A Short Video Sharing Mobile Application in China BT - Social
  Computing and Social Media. Design, Human Behavior and Analytics}}. In
  \bibinfo{booktitle}{\emph{International Conference on human-computer
  interaction}}, \bibfield{editor}{\bibinfo{person}{Gabriele Meiselwitz}}
  (Ed.). \bibinfo{publisher}{Springer International Publishing},
  \bibinfo{address}{Cham}, \bibinfo{pages}{233--244}.
\newblock
\showISBNx{978-3-030-21902-4}


\bibitem[\protect\citeauthoryear{Lu, Lu, and Liu}{Lu et~al\mbox{.}}{2020}]%
        {LuTikTokNonUse}
\bibfield{author}{\bibinfo{person}{Xing Lu}, \bibinfo{person}{Zhicong Lu},
  {and} \bibinfo{person}{Changqing Liu}.} \bibinfo{year}{2020}\natexlab{}.
\newblock \showarticletitle{{Exploring TikTok Use and Non-use Practices and
  Experiences in China}}. In \bibinfo{booktitle}{\emph{International conference
  on human-computer interaction}}, \bibfield{editor}{\bibinfo{person}{Gabriele
  Meiselwitz}} (Ed.). \bibinfo{publisher}{Springer International Publishing},
  \bibinfo{address}{Cham}, \bibinfo{pages}{57--70}.
\newblock
\showISBNx{978-3-030-49576-3}


\bibitem[\protect\citeauthoryear{Medina~Serrano, Papakyriakopoulos, and
  Hegelich}{Medina~Serrano et~al\mbox{.}}{2020}]%
        {SerranoTikTok}
\bibfield{author}{\bibinfo{person}{Juan~Carlos Medina~Serrano},
  \bibinfo{person}{Orestis Papakyriakopoulos}, {and} \bibinfo{person}{Simon
  Hegelich}.} \bibinfo{year}{2020}\natexlab{}.
\newblock \showarticletitle{{Dancing to the Partisan Beat: A First Analysis of
  Political Communication on TikTok}}. In \bibinfo{booktitle}{\emph{12th ACM
  Conference on Web Science}} \emph{(\bibinfo{series}{WebSci '20})}.
  \bibinfo{publisher}{Association for Computing Machinery},
  \bibinfo{address}{New York, NY, USA}, \bibinfo{pages}{257–266}.
\newblock
\showISBNx{9781450379892}
\urldef\tempurl%
\url{https://doi.org/10.1145/3394231.3397916}
\showDOI{\tempurl}


\bibitem[\protect\citeauthoryear{Meng and Leung}{Meng and Leung}{2021}]%
        {MengTikTok}
\bibfield{author}{\bibinfo{person}{Keira~Shuyang Meng} {and}
  \bibinfo{person}{Louis Leung}.} \bibinfo{year}{2021}\natexlab{}.
\newblock \showarticletitle{{Factors influencing TikTok engagement behaviors in
  China: An examination of gratifications sought, narcissism, and the Big Five
  personality traits}}.
\newblock \bibinfo{journal}{\emph{Telecommunications Policy}}
  \bibinfo{volume}{45}, \bibinfo{number}{7} (\bibinfo{year}{2021}),
  \bibinfo{pages}{102172}.
\newblock
\showISSN{0308-5961}
\urldef\tempurl%
\url{https://doi.org/10.1016/j.telpol.2021.102172}
\showDOI{\tempurl}


\bibitem[\protect\citeauthoryear{Munger and Phillips}{Munger and
  Phillips}{2019}]%
        {MungerSupplyDemand}
\bibfield{author}{\bibinfo{person}{Kevin Munger} {and} \bibinfo{person}{Joseph
  Phillips}.} \bibinfo{year}{2019}\natexlab{}.
\newblock \showarticletitle{{A Supply and Demand Framework for YouTube
  Politics}}.
\newblock \bibinfo{journal}{\emph{Unpublished Paper}} (\bibinfo{year}{2019}).
\newblock


\bibitem[\protect\citeauthoryear{Niu, Mai, McKim, and McCrickard}{Niu
  et~al\mbox{.}}{2021}]%
        {NiuTeamTrees}
\bibfield{author}{\bibinfo{person}{Shuo Niu}, \bibinfo{person}{Cat Mai},
  \bibinfo{person}{Katherine~G McKim}, {and} \bibinfo{person}{D~Scott
  McCrickard}.} \bibinfo{year}{2021}\natexlab{}.
\newblock \showarticletitle{{{\#}TeamTrees: Investigating How YouTubers
  Participate in a Social Media Campaign}}.
\newblock \bibinfo{journal}{\emph{Proc. ACM Hum.-Comput. Interact.}}
  \bibinfo{volume}{5}, \bibinfo{number}{CSCW2} (\bibinfo{date}{10}
  \bibinfo{year}{2021}).
\newblock
\urldef\tempurl%
\url{https://doi.org/10.1145/3479593}
\showDOI{\tempurl}


\bibitem[\protect\citeauthoryear{Omar and Dequan}{Omar and Dequan}{2020}]%
        {OmarTikTok}
\bibfield{author}{\bibinfo{person}{Bahiyah Omar} {and} \bibinfo{person}{Wang
  Dequan}.} \bibinfo{year}{2020}\natexlab{}.
\newblock \bibinfo{title}{{Watch, Share or Create: The Influence of Personality
  Traits and User Motivation on TikTok Mobile Video Usage}}.
\newblock
\newblock
\urldef\tempurl%
\url{https://www.learntechlib.org/p/216454}
\showURL{%
\tempurl}


\bibitem[\protect\citeauthoryear{Ottoni, Cunha, Magno, Bernardina, Meira~Jr.,
  and Almeida}{Ottoni et~al\mbox{.}}{2018}]%
        {OttoniRightWingYouTubeChannels}
\bibfield{author}{\bibinfo{person}{Raphael Ottoni}, \bibinfo{person}{Evandro
  Cunha}, \bibinfo{person}{Gabriel Magno}, \bibinfo{person}{Pedro Bernardina},
  \bibinfo{person}{Wagner Meira~Jr.}, {and} \bibinfo{person}{Virg\'{\i}lio
  Virgílio Virg\'{\i}lio~Virgílio Almeida}.} \bibinfo{year}{2018}\natexlab{}.
\newblock \showarticletitle{{Analyzing Right-Wing YouTube Channels: Hate,
  Violence and Discrimination}}. In \bibinfo{booktitle}{\emph{Proceedings of
  the 10th ACM Conference on Web Science}} \emph{(\bibinfo{series}{WebSci
  '18})}. \bibinfo{publisher}{Association for Computing Machinery},
  \bibinfo{address}{New York, NY, USA}, \bibinfo{pages}{323–332}.
\newblock
\showISBNx{9781450355636}
\urldef\tempurl%
\url{https://doi.org/10.1145/3201064.3201081}
\showDOI{\tempurl}


\bibitem[\protect\citeauthoryear{Perrin and Anderson}{Perrin and
  Anderson}{2019}]%
        {perrin_anderson_2019}
\bibfield{author}{\bibinfo{person}{Andrew Perrin} {and} \bibinfo{person}{Monica
  Anderson}.} \bibinfo{year}{2019}\natexlab{}.
\newblock \bibinfo{title}{{Share of U.S. adults using social media, including
  Facebook, is mostly unchanged since 2018}}.
\newblock
\newblock


\bibitem[\protect\citeauthoryear{Raby, Caron, Th{\'{e}}wissen-LeBlanc,
  Prioletta, and Mitchell}{Raby et~al\mbox{.}}{2018}]%
        {RabyYouTubeSocialChange}
\bibfield{author}{\bibinfo{person}{Rebecca Raby}, \bibinfo{person}{Caroline
  Caron}, \bibinfo{person}{Sophie Th{\'{e}}wissen-LeBlanc},
  \bibinfo{person}{Jessica Prioletta}, {and} \bibinfo{person}{Claudia
  Mitchell}.} \bibinfo{year}{2018}\natexlab{}.
\newblock \showarticletitle{{Vlogging on YouTube: the online, political
  engagement of young Canadians advocating for social change}}.
\newblock \bibinfo{journal}{\emph{Journal of Youth Studies}}
  \bibinfo{volume}{21}, \bibinfo{number}{4} (\bibinfo{date}{4}
  \bibinfo{year}{2018}), \bibinfo{pages}{495--512}.
\newblock
\showISSN{1367-6261}
\urldef\tempurl%
\url{https://doi.org/10.1080/13676261.2017.1394995}
\showDOI{\tempurl}


\bibitem[\protect\citeauthoryear{Rappoport and Kren}{Rappoport and
  Kren}{1975}]%
        {RappoportSocialIssue}
\bibfield{author}{\bibinfo{person}{Leon Rappoport} {and}
  \bibinfo{person}{George Kren}.} \bibinfo{year}{1975}\natexlab{}.
\newblock \showarticletitle{{What is a social issue?}}
\newblock \bibinfo{journal}{\emph{American Psychologist}} \bibinfo{volume}{30},
  \bibinfo{number}{8} (\bibinfo{date}{8} \bibinfo{year}{1975}),
  \bibinfo{pages}{838--841}.
\newblock
\showISSN{0003-066X}
\urldef\tempurl%
\url{https://doi.org/10.1037/h0077076}
\showDOI{\tempurl}


\bibitem[\protect\citeauthoryear{Rohde, Aal, Misaki, Randall, Weibert, and
  Wulf}{Rohde et~al\mbox{.}}{2016}]%
        {RohdeSyriaVideo}
\bibfield{author}{\bibinfo{person}{Markus Rohde}, \bibinfo{person}{Konstantin
  Aal}, \bibinfo{person}{Kaoru Misaki}, \bibinfo{person}{Dave Randall},
  \bibinfo{person}{Anne Weibert}, {and} \bibinfo{person}{Volker Wulf}.}
  \bibinfo{year}{2016}\natexlab{}.
\newblock \showarticletitle{{Out of Syria: Mobile Media in Use at the Time of
  Civil War}}.
\newblock \bibinfo{journal}{\emph{International Journal of Human–Computer
  Interaction}} \bibinfo{volume}{32}, \bibinfo{number}{7} (\bibinfo{date}{7}
  \bibinfo{year}{2016}), \bibinfo{pages}{515--531}.
\newblock
\showISSN{1044-7318}
\urldef\tempurl%
\url{https://doi.org/10.1080/10447318.2016.1177300}
\showDOI{\tempurl}


\bibitem[\protect\citeauthoryear{Rotman and Preece}{Rotman and Preece}{2010}]%
        {PreeceWeTube}
\bibfield{author}{\bibinfo{person}{Dana Rotman} {and} \bibinfo{person}{Jennifer
  Preece}.} \bibinfo{year}{2010}\natexlab{}.
\newblock \showarticletitle{{The'WeTube'in YouTube–creating an online
  community through video sharing}}.
\newblock \bibinfo{journal}{\emph{International Journal of Web Based
  Communities}} \bibinfo{volume}{6}, \bibinfo{number}{3}
  (\bibinfo{year}{2010}), \bibinfo{pages}{317--333}.
\newblock
\showISSN{1477-8394}


\bibitem[\protect\citeauthoryear{Shahid, Kamath, Sidotam, Jiang, Batino, and
  Vashistha}{Shahid et~al\mbox{.}}{2022}]%
        {ShahidFakeVideos}
\bibfield{author}{\bibinfo{person}{Farhana Shahid}, \bibinfo{person}{Srujana
  Kamath}, \bibinfo{person}{Annie Sidotam}, \bibinfo{person}{Vivian Jiang},
  \bibinfo{person}{Alexa Batino}, {and} \bibinfo{person}{Aditya Vashistha}.}
  \bibinfo{year}{2022}\natexlab{}.
\newblock \showarticletitle{{”It Matches My Worldview”: Examining
  Perceptions and Attitudes Around Fake Videos}}. In
  \bibinfo{booktitle}{\emph{CHI Conference on Human Factors in Computing
  Systems}} \emph{(\bibinfo{series}{CHI '22})}. \bibinfo{publisher}{Association
  for Computing Machinery}, \bibinfo{address}{New York, NY, USA}.
\newblock
\showISBNx{9781450391573}
\urldef\tempurl%
\url{https://doi.org/10.1145/3491102.3517646}
\showDOI{\tempurl}


\bibitem[\protect\citeauthoryear{Shapiro and Park}{Shapiro and Park}{2014}]%
        {Shapiro2014}
\bibfield{author}{\bibinfo{person}{Matthew~A Shapiro} {and}
  \bibinfo{person}{Han~Woo Park}.} \bibinfo{year}{2014}\natexlab{}.
\newblock \showarticletitle{{More than entertainment: YouTube and public
  responses to the science of global warming and climate change}}.
\newblock \bibinfo{journal}{\emph{Social Science Information}}
  \bibinfo{volume}{54}, \bibinfo{number}{1} (\bibinfo{date}{11}
  \bibinfo{year}{2014}), \bibinfo{pages}{115--145}.
\newblock
\showISSN{0539-0184}
\urldef\tempurl%
\url{https://doi.org/10.1177/0539018414554730}
\showDOI{\tempurl}


\bibitem[\protect\citeauthoryear{Shaw and Hargittai}{Shaw and
  Hargittai}{2021}]%
        {ShawMTurk}
\bibfield{author}{\bibinfo{person}{Aaron Shaw} {and} \bibinfo{person}{Eszter
  Hargittai}.} \bibinfo{year}{2021}\natexlab{}.
\newblock \showarticletitle{{Do the online activities of Amazon Mechanical Turk
  workers mirror those of the general population?: A comparison of two survey
  samples}}.
\newblock \bibinfo{journal}{\emph{International Journal of Communication}}
  \bibinfo{volume}{15} (\bibinfo{year}{2021}), \bibinfo{pages}{4383--4398}.
\newblock
\showISSN{1932-8036}
\urldef\tempurl%
\url{https://www.zora.uzh.ch/id/eprint/210738/}
\showURL{%
\tempurl}


\bibitem[\protect\citeauthoryear{Shutsko}{Shutsko}{2020}]%
        {Shutsko2020TikTok}
\bibfield{author}{\bibinfo{person}{Aliaksandra Shutsko}.}
  \bibinfo{year}{2020}\natexlab{}.
\newblock \showarticletitle{{User-Generated Short Video Content in Social
  Media. A Case Study of TikTok}}, \bibfield{editor}{\bibinfo{person}{Gabriele
  Meiselwitz}} (Ed.). \bibinfo{publisher}{Springer International Publishing},
  \bibinfo{address}{Cham}, \bibinfo{pages}{108--125}.
\newblock
\showISBNx{978-3-030-49576-3}


\bibitem[\protect\citeauthoryear{Stocking, Kessel, Barthel, Matsa, and
  Khuzam}{Stocking et~al\mbox{.}}{2020}]%
        {StockingPewResearch}
\bibfield{author}{\bibinfo{person}{Galen Stocking},
  \bibinfo{person}{Patrick~van Kessel}, \bibinfo{person}{Michael Barthel},
  \bibinfo{person}{Katerina~Eva Matsa}, {and} \bibinfo{person}{Maya Khuzam}.}
  \bibinfo{year}{2020}\natexlab{}.
\newblock \showarticletitle{{Many Americans Get News on YouTube, Where News
  Organizations and Independent Producers Thrive Side by Side}}.
\newblock \bibinfo{journal}{\emph{Pew Research Center's Journalism Project}}
  (\bibinfo{date}{5} \bibinfo{year}{2020}).
\newblock


\bibitem[\protect\citeauthoryear{Tang, Fujimoto, Amith, Cunningham, Costantini,
  York, Xiong, Boom, and Tao}{Tang et~al\mbox{.}}{2021}]%
        {TangMisinformation}
\bibfield{author}{\bibinfo{person}{Lu Tang}, \bibinfo{person}{Kayo Fujimoto},
  \bibinfo{person}{Muhammad~(Tuan) Amith}, \bibinfo{person}{Rachel Cunningham},
  \bibinfo{person}{Rebecca~A Costantini}, \bibinfo{person}{Felicia York},
  \bibinfo{person}{Grace Xiong}, \bibinfo{person}{Julie~A Boom}, {and}
  \bibinfo{person}{Cui Tao}.} \bibinfo{year}{2021}\natexlab{}.
\newblock \showarticletitle{{“Down the Rabbit Hole” of Vaccine
  Misinformation on YouTube: Network Exposure Study}}.
\newblock \bibinfo{journal}{\emph{J Med Internet Res}} \bibinfo{volume}{23},
  \bibinfo{number}{1} (\bibinfo{year}{2021}), \bibinfo{pages}{e23262}.
\newblock
\showISSN{1438-8871}
\urldef\tempurl%
\url{https://doi.org/10.2196/23262}
\showDOI{\tempurl}


\bibitem[\protect\citeauthoryear{Towner and Dulio}{Towner and Dulio}{2011}]%
        {TownerYouTubeElection}
\bibfield{author}{\bibinfo{person}{Terri~L Towner} {and}
  \bibinfo{person}{David~A Dulio}.} \bibinfo{year}{2011}\natexlab{}.
\newblock \showarticletitle{{An experiment of campaign effects during the
  YouTube election}}.
\newblock \bibinfo{journal}{\emph{New Media {\&} Society}}
  \bibinfo{volume}{13}, \bibinfo{number}{4} (\bibinfo{date}{2}
  \bibinfo{year}{2011}), \bibinfo{pages}{626--644}.
\newblock
\showISSN{1461-4448}
\urldef\tempurl%
\url{https://doi.org/10.1177/1461444810377917}
\showDOI{\tempurl}


\bibitem[\protect\citeauthoryear{Vijay and Gekker}{Vijay and Gekker}{2021}]%
        {VijayTikTokPolitics}
\bibfield{author}{\bibinfo{person}{Darsana Vijay} {and} \bibinfo{person}{Alex
  Gekker}.} \bibinfo{year}{2021}\natexlab{}.
\newblock \showarticletitle{{Playing Politics: How Sabarimala Played Out on
  TikTok}}.
\newblock \bibinfo{journal}{\emph{American Behavioral Scientist}}
  \bibinfo{volume}{65}, \bibinfo{number}{5} (\bibinfo{date}{1}
  \bibinfo{year}{2021}), \bibinfo{pages}{712--734}.
\newblock
\showISSN{0002-7642}
\urldef\tempurl%
\url{https://doi.org/10.1177/0002764221989769}
\showDOI{\tempurl}


\bibitem[\protect\citeauthoryear{Wang}{Wang}{2014}]%
        {WangUGTYouTube}
\bibfield{author}{\bibinfo{person}{Tai-Li Wang}.}
  \bibinfo{year}{2014}\natexlab{}.
\newblock \showarticletitle{{The usage behaviors, motivations and
  gratifications of using User-Generated Media: The case study of Taiwan’s
  YouTube}}.
\newblock \bibinfo{journal}{\emph{Advances in Journalism and Communication}}
  \bibinfo{volume}{2}, \bibinfo{number}{04} (\bibinfo{year}{2014}),
  \bibinfo{pages}{137}.
\newblock


\bibitem[\protect\citeauthoryear{Weimann and Masri}{Weimann and Masri}{2020}]%
        {WeimannSpreadingHateTikTok}
\bibfield{author}{\bibinfo{person}{Gabriel Weimann} {and}
  \bibinfo{person}{Natalie Masri}.} \bibinfo{year}{2020}\natexlab{}.
\newblock \showarticletitle{{Research Note: Spreading Hate on TikTok}}.
\newblock \bibinfo{journal}{\emph{Studies in Conflict {\&} Terrorism}}
  (\bibinfo{date}{6} \bibinfo{year}{2020}), \bibinfo{pages}{1--14}.
\newblock
\showISSN{1057-610X}
\urldef\tempurl%
\url{https://doi.org/10.1080/1057610X.2020.1780027}
\showDOI{\tempurl}


\bibitem[\protect\citeauthoryear{Whiting and Williams}{Whiting and
  Williams}{2013}]%
        {WhitingUseGratification}
\bibfield{author}{\bibinfo{person}{Anita Whiting} {and} \bibinfo{person}{David
  Williams}.} \bibinfo{year}{2013}\natexlab{}.
\newblock \showarticletitle{{Why people use social media: a uses and
  gratifications approach}}.
\newblock \bibinfo{journal}{\emph{Qualitative Market Research: An International
  Journal}} \bibinfo{volume}{16}, \bibinfo{number}{4} (\bibinfo{date}{1}
  \bibinfo{year}{2013}), \bibinfo{pages}{362--369}.
\newblock
\showISSN{1352-2752}
\urldef\tempurl%
\url{https://doi.org/10.1108/QMR-06-2013-0041}
\showDOI{\tempurl}


\bibitem[\protect\citeauthoryear{Wohn, Fiesler, Hemphill, De~Choudhury, and
  Matias}{Wohn et~al\mbox{.}}{2017}]%
        {WohnContentCuration}
\bibfield{author}{\bibinfo{person}{Donghee~Yvette Wohn}, \bibinfo{person}{Casey
  Fiesler}, \bibinfo{person}{Libby Hemphill}, \bibinfo{person}{Munmun
  De~Choudhury}, {and} \bibinfo{person}{J~Nathan Matias}.}
  \bibinfo{year}{2017}\natexlab{}.
\newblock \showarticletitle{{How to Handle Online Risks? Discussing Content
  Curation and Moderation in Social Media}}. In
  \bibinfo{booktitle}{\emph{Proceedings of the 2017 CHI Conference Extended
  Abstracts on Human Factors in Computing Systems}} \emph{(\bibinfo{series}{CHI
  EA '17})}. \bibinfo{publisher}{Association for Computing Machinery},
  \bibinfo{address}{New York, NY, USA}, \bibinfo{pages}{1271–1276}.
\newblock
\showISBNx{9781450346566}
\urldef\tempurl%
\url{https://doi.org/10.1145/3027063.3051141}
\showDOI{\tempurl}


\bibitem[\protect\citeauthoryear{Wohn, Freeman, and McLaughlin}{Wohn
  et~al\mbox{.}}{2018}]%
        {WohnParasocialInteraction}
\bibfield{author}{\bibinfo{person}{Donghee~Yvette Wohn}, \bibinfo{person}{Guo
  Freeman}, {and} \bibinfo{person}{Caitlin McLaughlin}.}
  \bibinfo{year}{2018}\natexlab{}.
\newblock \showarticletitle{{Explaining Viewers' Emotional, Instrumental, and
  Financial Support Provision for Live Streamers}}. In
  \bibinfo{booktitle}{\emph{Proceedings of the 2018 CHI Conference on Human
  Factors in Computing Systems}} \emph{(\bibinfo{series}{CHI '18})}.
  \bibinfo{publisher}{Association for Computing Machinery},
  \bibinfo{address}{New York, NY, USA}, \bibinfo{pages}{1–13}.
\newblock
\showISBNx{9781450356206}
\urldef\tempurl%
\url{https://doi.org/10.1145/3173574.3174048}
\showDOI{\tempurl}


\bibitem[\protect\citeauthoryear{Yang and Ha}{Yang and Ha}{2021}]%
        {YangTikTokChina}
\bibfield{author}{\bibinfo{person}{Yang Yang} {and} \bibinfo{person}{Louisa
  Ha}.} \bibinfo{year}{2021}\natexlab{}.
\newblock \showarticletitle{{Why People Use TikTok (Douyin) and How Their
  Purchase Intentions Are Affected by Social Media Influencers in China: A Uses
  and Gratifications and Parasocial Relationship Perspective}}.
\newblock \bibinfo{journal}{\emph{Journal of Interactive Advertising}}
  \bibinfo{volume}{21}, \bibinfo{number}{3} (\bibinfo{date}{9}
  \bibinfo{year}{2021}), \bibinfo{pages}{297--305}.
\newblock
\showISSN{null}
\urldef\tempurl%
\url{https://doi.org/10.1080/15252019.2021.1995544}
\showDOI{\tempurl}


\bibitem[\protect\citeauthoryear{Yaqub, Kakhidze, Brockman, Memon, and
  Patil}{Yaqub et~al\mbox{.}}{2020}]%
        {PatilCredibilityIndicator}
\bibfield{author}{\bibinfo{person}{Waheeb Yaqub}, \bibinfo{person}{Otari
  Kakhidze}, \bibinfo{person}{Morgan~L Brockman}, \bibinfo{person}{Nasir
  Memon}, {and} \bibinfo{person}{Sameer Patil}.}
  \bibinfo{year}{2020}\natexlab{}.
\newblock \showarticletitle{{Effects of Credibility Indicators on Social Media
  News Sharing Intent}}.
\newblock In \bibinfo{booktitle}{\emph{Proceedings of the 2020 CHI Conference
  on Human Factors in Computing Systems}}. \bibinfo{publisher}{Association for
  Computing Machinery}, \bibinfo{address}{New York, NY, USA},
  \bibinfo{pages}{1–14}.
\newblock
\showISBNx{9781450367080}
\urldef\tempurl%
\url{https://doi.org/10.1145/3313831.3376213}
\showURL{%
\tempurl}


\bibitem[\protect\citeauthoryear{Zhang and Liu}{Zhang and Liu}{2021}]%
        {Zhang2021TikTokAlgorithm}
\bibfield{author}{\bibinfo{person}{Min Zhang} {and} \bibinfo{person}{Yiqun
  Liu}.} \bibinfo{year}{2021}\natexlab{}.
\newblock \showarticletitle{{A commentary of TikTok recommendation algorithms
  in MIT Technology Review 2021}}.
\newblock \bibinfo{journal}{\emph{Fundamental Research}} \bibinfo{volume}{1},
  \bibinfo{number}{6} (\bibinfo{year}{2021}), \bibinfo{pages}{846--847}.
\newblock
\showISSN{2667-3258}
\urldef\tempurl%
\url{https://doi.org/10.1016/j.fmre.2021.11.015}
\showDOI{\tempurl}


\bibitem[\protect\citeauthoryear{Zhang}{Zhang}{2021}]%
        {Zhang2021TikTok}
\bibfield{author}{\bibinfo{person}{Zongyi Zhang}.}
  \bibinfo{year}{2021}\natexlab{}.
\newblock \showarticletitle{{Infrastructuralization of Tik Tok: Transformation,
  power relationships, and platformization of video entertainment in China}}.
\newblock \bibinfo{journal}{\emph{Media, Culture {\&} Society}}
  \bibinfo{volume}{43}, \bibinfo{number}{2} (\bibinfo{year}{2021}),
  \bibinfo{pages}{219--236}.
\newblock
\showISSN{0163-4437}


\bibitem[\protect\citeauthoryear{Zimmermann, Noll, Gr{\"{a}}{\ss}er, Hugger,
  Braun, Nowak, and Kaspar}{Zimmermann et~al\mbox{.}}{2020}]%
        {ZimmermannYouTube}
\bibfield{author}{\bibinfo{person}{Daniel Zimmermann},
  \bibinfo{person}{Christian Noll}, \bibinfo{person}{Lars Gr{\"{a}}{\ss}er},
  \bibinfo{person}{Kai-Uwe Hugger}, \bibinfo{person}{Lea~Marie Braun},
  \bibinfo{person}{Tine Nowak}, {and} \bibinfo{person}{Kai Kaspar}.}
  \bibinfo{year}{2020}\natexlab{}.
\newblock \showarticletitle{{Influencers on YouTube: a quantitative study on
  young people’s use and perception of videos about political and societal
  topics}}.
\newblock \bibinfo{journal}{\emph{Current Psychology}} (\bibinfo{year}{2020}).
\newblock
\showISSN{1936-4733}
\urldef\tempurl%
\url{https://doi.org/10.1007/s12144-020-01164-7}
\showDOI{\tempurl}


\end{thebibliography}

\end{document}